\documentclass[aps,prl,twocolumn,showpacs,superscriptaddress]{revtex4-1}
\usepackage{graphicx}
\usepackage{amsmath}
\usepackage{amssymb}
\usepackage{epstopdf}
\usepackage{amstext}
\usepackage{amsthm}
\usepackage{amsfonts}
\usepackage{color}
\usepackage[normalem]{ulem}
 
\usepackage[unicode]{hyperref}
\hypersetup{
	pdftitle={soliton types},
	pdfauthor={},
	pdfsubject={},
	colorlinks=true,
	linkcolor=blue,
	citecolor=blue,
	filecolor=black,
	urlcolor=blue
}
\def\urlprefix{}
\def\url#1{}
\usepackage{url}

\usepackage{bm}
\usepackage{color}
\begin{document}
	
	
	\title{Two types of dark solitons  in a spin-orbit-coupled Fermi gas}
	
	\author{Antonio Mu\~noz Mateo}\email{ammateo67@gmail.com}
	\affiliation{
		Departamento de F\'{i}sica, Facultad de Ciencias, Universidad de La 
		Laguna, E¡§C-38200 La Laguna, Tenerife, Spain
	}
	\affiliation{Graduate School of  China Academy of Engineering Physics, Beijing 100193, China}
	\author{Xiaoquan Yu }\email{xqyu@gscaep.ac.cn}
	\affiliation{Graduate School of  China Academy of Engineering Physics, Beijing 100193, China}
	\affiliation{Department of Physics, Centre for Quantum Science, and Dodd-Walls Centre for Photonic and Quantum Technologies, University of Otago, Dunedin 9016, New Zealand}

	\begin{abstract}
		Dark solitons in quantum fluids are well known nonlinear excitations that are usually
		characterized by a single length scale associated with the underlying background fluid. We show that in the presence of spin-orbit coupling and a linear Zeeman field, superfluid Fermi gases support two different types of nonlinear excitations featured by corresponding length scales related to the existence of two Fermi surfaces. Only one of these types, which occurs for finite spin-orbit coupling and a Zeeman field, survives to the topological phase transition, and is therefore capable to sustain Majorana zero modes. At the point of the emergence of this soliton for varying \sout{the} Zeeman field, the associated Andreev bound states present a minigap that vanishes for practical purposes, in spite of lacking the reality condition of Majorana modes.
	\end{abstract}
	
	\maketitle

	\textit{Introduction.---}
	Dark solitons are topological excitations that result from the balance between interaction and kinetic energy~\cite{Frantzeskakis2010}. In ultra-cold Fermi gases~\cite{GiorginiRMP2008}, a dark soliton is a phase domain wall in the pairing wave function (or order parameter), which vanishes at the soliton core and shows a $\pi$  phase jump across it. Dark solitons probe features of the underlying superfluidity of the Fermi gas,
	and provide a connection between macroscopic motion and dynamics at the interatomic length scale.
	
	The static structure, dynamics and stability of dark solitons in ordinary Fermi gases have been widely investigated, both  theoretically~\cite{Antezza2007,Scott2011,Liao2011,Spuntarelli2011,Cetoli2013,Efimkin2015} and experimentally~\cite{Yefsah2013,Ku2014,Zwierleinprl2014}. Meanwhile, the properties of solitons in spin-orbit (SO) coupled Fermi gases~\cite{Wang2012, Cheuk2012, Fu2013,Williams2013,Fu2014,Burdick2016,Song2016,Zhai2015} are less well understood.  
	In the presence of a SO coupling and a linear Zeeman field, an interacting Fermi gas exhibits a topological phase transition between the regular superfluid phase and the topological superfluid phase, where the latter one supports Majorana zero modes (MZMs)~\cite{Oreg2010,Lutchyn2010, Wei2012,LiuHu2012,Liu2012}. The MZMs can be found when the fermionic pairing vanishes locally, and thus they are associated either with the system boundary, as edge states, or with internal, local defects which locally destroy superfluidity, as pinned modes. 
	A particularly interesting example of the latter in one dimension (1D) is the dark soliton. MZMs have striking features~\cite{Kitaev2001, Alicea2012,aguado2017majorana} and have potential application in fault-tolerant quantum computation~\cite{kitaev2003fault,sarma2015majorana,rmptqc}. In addition, dark solitons hosting MZMs exhibit novel dynamics distinct from the normal behavior of solitons~\cite{Zou2016}.  
	In SO coupled Fermi gases, most of the attention has been focused on solitons that smoothly connect to ordinary solitons when the SO coupling and the Zeeman field go to zero~\cite{Xu2014,Liu2015}. The presence in this system of two Fermi surfaces~\cite{Thompson2020}, with different characteristic energy and length scales that feature distinct condensation peaks of fermionic pairs, suggests the possible existence of different types of topological excitations for finite SO couplings.  However, to the best of our knowledge,  this possibility has been overlooked.
	
	
	In this work we show that in the presence of SO coupling and Zeeman field, the Fermi gas supports two different types of dark solitons characterized by length scales related to the existence of two, inner and outer, Fermi surfaces. In complement to previous studies~\cite{Xu2014,Liu2015}, we find that (i) a new type of soliton associated with the outer Fermi surface, existing only in the presence of SO coupling and a Zeeman field, has continuation (as based on the continuous existence of such Fermi surface) into the topological regime where it hosts MZMs at the core, (ii) the onset of this soliton is accompanied by the appearance of non-topological quasi-zero-energy Andreev bound states (ABSs) inside the core, (iii) the soliton associated with the inner Fermi surface, which smoothly connects to the regular soliton without SO coupling, has no continuation into the topological regime as its characteristic length scale vanishes when approaching the transition point, and (iv) the order parameter profile, the particle density, and the associated ABS spectrum are distinct for the two types of solitons. This characterization also allows us to propose accurate ansatzes to describe MZMs inside the soliton core.

	\textit{Model.---}
	We consider a 1D spin-1/2 Fermi gas with SO coupling at zero temperature. Within a mean field approach,  the energy spectrum $E_j$ and the corresponding fermionic quasi-particle amplitudes $\{u_{\sigma j}(x),v_{\sigma j}(x)\}$ with spin $\sigma=\uparrow,\downarrow$ are given by the Bogoliubov-de Gennes (BdG) equations~\cite{Liu2012,LiuHu2012,Xu2014,Liu2015,footnoteMZM}  
	\begin{align}
	\begin{bmatrix}
	\hat H_{so} && i\Delta\sigma_y \\
	(i\Delta\sigma_y)^\dagger && -\sigma_x\hat H_{so}\sigma_x
	\end{bmatrix} \,\psi_j
	=E_{j} \psi_j ,
	\label{eq:BdG}
	\end{align}
	where $\psi_j=[u_{\uparrow j}, \, u_{\downarrow j},\, v_{\uparrow j},\, v_{\downarrow j}]^T$ and $j=1,2,\dots$ labels the state, and the single-particle Hamiltonian is
	\begin{align}
	\hat H_{so}=-\frac{\hbar^2}{2m}\partial_x^2+ V_{\rm ext}(x)-\mu_\sigma + \frac{\hbar k_\ell}{m} \hat{p}_x \,\sigma_z -\nu\, \sigma_x.
	\end{align}
	Here $\sigma_{i=x,y,z}$ are Pauli matrices, $\nu$ denotes the strength of the Zeeman field (or linear coupling), 
	$k_l$ couples the orbit and spin degrees of freedom, and $ V_{\rm ext}(x)$ is the confining potential. We focus on  spin-balanced systems~\cite{footnoteMZM}, with chemical potential  $\mu_\uparrow=\mu_\downarrow=\mu$. The BdG equation~\eqref{eq:BdG} has the particle-hole
	symmetry, i.e., $C \psi^{*}_{E_j}= \psi_{-E_j}$ that connects the positive and negative energy states through $[u_{\sigma},\,v_{\sigma}]\rightarrow e^{i \phi}[v^{*}_{\sigma},\,u^{*}_{\sigma}]$ as $E_j \rightarrow -E_j$, where $C$ satisfies $C^{*}C=I$~\cite{footnoteC}. Hence the two eigenstates corresponding to energies $\pm E_j$ describe the same physical degrees of freedom. 
	The modes that satisfy the reality condition $C \psi^{*}_{E_j}=\psi_{E_j}$  are Majorana Fermions~\cite{Chamon2010,Jackiw_2012,aguado2017majorana}.   The particle-hole symmetry ensures that the reality condition can be achieved only for $E_j=0$, i.e., $C \psi^{*}_{0}= \psi_{0}$ or $u_{\sigma}= e^{i \phi}v^{*}_{\sigma}$.
	At zero temperature, the number density can be written as
	$n(x)=\sum_{j,\sigma, E_j\geq0}|v_{\sigma,j}(x)|^2$,
	and the order parameter of paired fermions as
	$\Delta (x) = {g_{\rm 1D}}\sum_{j,E_j\geq0} u_{\downarrow j}(x)\,v_{\uparrow j}^*(x) $,
	where $g_{\rm 1D}<0$ is the 1D attractive interaction strength between opposite spin particles.  We characterize the interaction by the non-dimensional parameter
	$\gamma={m\,|g_{1D}|}/({\pi\,\hbar^2\,k_{TF}})$,
	where $k_{TF}=\pi n_{TF}/2$ and $n_{TF}$ are the Fermi wavenumber and 
	the number density, respectively, of the noninteracting gas.

	\textit{Two Fermi surfaces.---}
	For the static, uniform density state, the plane-wave expansion of the spinor $\psi_k(x)=[u_{\uparrow k},\, u_{\downarrow k},\,v_{\uparrow k},\, v_{\downarrow k}]^T\,\exp(ikx)/\sqrt{2\pi}$ provides the dispersion (positive energy branches)
	\begin{align}
	E_{1,2}(k)=\sqrt{\varepsilon_k^2+\nu^2+\zeta_\ell ^2 \pm \,2 \sqrt{\zeta_k^2\, \zeta_\ell  ^2+\varepsilon_k^2\,\nu^2}}
	\label{eq:FGso_disp}
	\end{align}
	where $\zeta_k =\hbar^2 k^2 \,/(2m)-\mu$, $\zeta_\ell =\hbar^2 k_\ell \, k/m$, $\varepsilon_k=\sqrt{\zeta_k^2+|\Delta|^2}$ is the eigen-energy of the Fermi gas in the absence of SO coupling and $E_2(k)>E_1(k)$. For $k=0$, Eq.(\ref{eq:FGso_disp}) gives $\hbar\omega_0=\nu\pm\sqrt{\mu^2+|\Delta|^2}$.  The energy gap of the lower branch is closed for $\nu_c=\sqrt{\mu^2+|\Delta|^2}$. For $\nu>\nu_c$ the gap reopens, and the system enters the topological regime \cite{Lutchyn2010, Oreg2010,LiuHu2012,Wei2012}. 
	Such closing and re-opening of the energy gap is an instance
	of a topological transition: broadly speaking, 
	a transition that separates two phases
	characterised by the value of a topological invariant (instead of a broken symmetry)~\cite{Alicea2012}.

	\begin{figure}[t]
		\centering
		\includegraphics[width=1.0\linewidth]{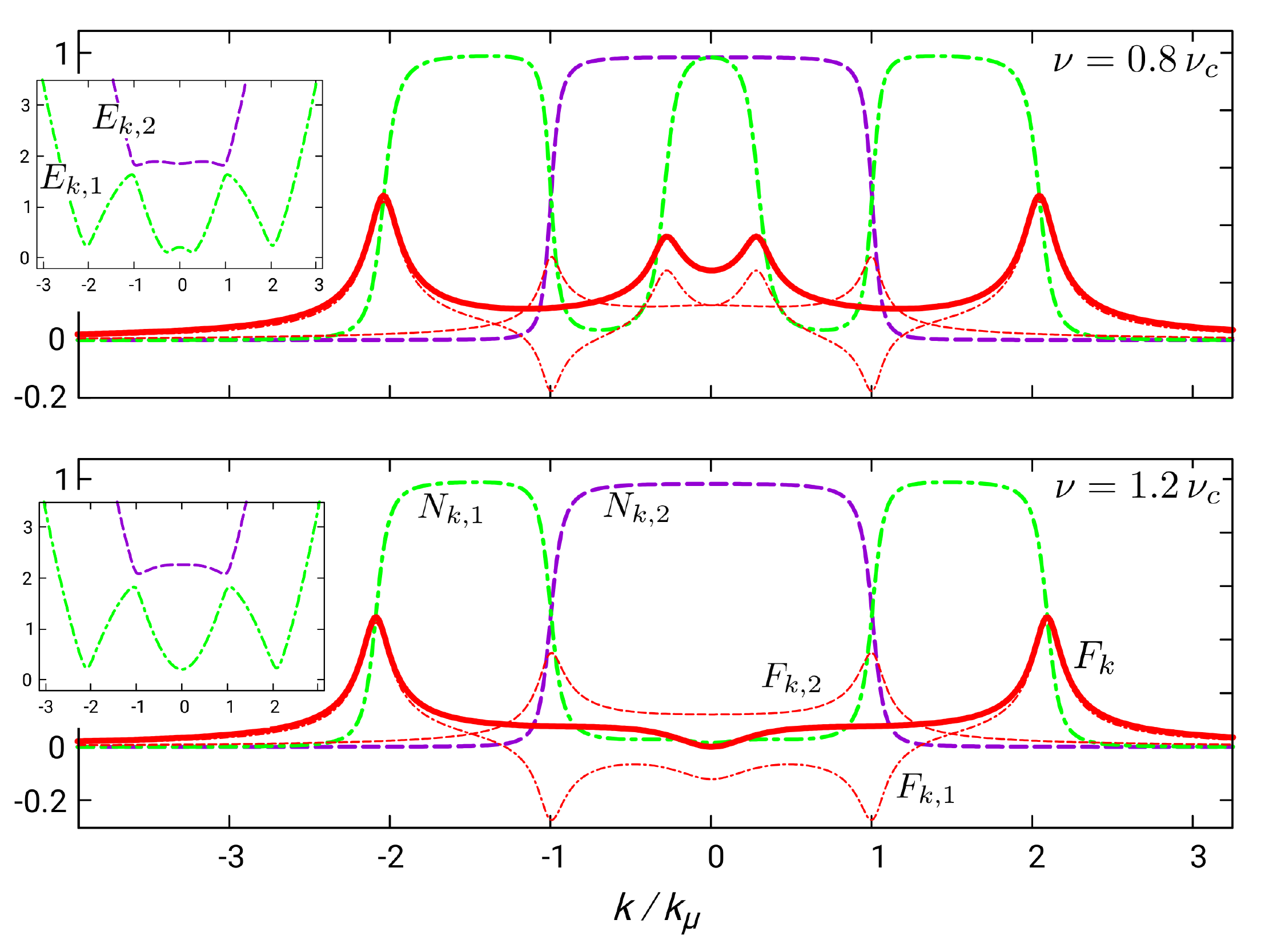}
		\caption{Momentum distribution $N_k=\sum_\sigma v_{\sigma k}^2=N_{k,1}+N_{k,2}$(dashed line) and symmetric condensation amplitude $F_k= u_{\downarrow k}v_{\uparrow k}^*-u_{\uparrow k}v_{\downarrow k}^*=F_{k,1}+F_{k,2}$ (solid line) in the regular superfluid phase (top panel) and in the topological phase (bottom panel), where $N_{k,i=1,2}$ and $F_{k,i=1,2}$ account for the contributions from each band. The condensation amplitude peaks at the position of the Fermi surfaces. Here $k_l=0.75 k_{\mu}$ and  $\Delta=0.25\,\mu$.  The inset shows the dispersion of the two positive-energy bands $E_{1,2}(k)$.  }
		\label{fig:FGso}
	\end{figure}
	Particular features introduced by the SO coupling emerge from the two-band structure of the dispersion. These bands give rise to two Fermi surfaces  associated with Fermi wave vectors $k_{F \pm}=\pi n_\pm/2$, where $n_{\pm}$ represent different contributions to the total number density, $n=n_+ +n_-$, from both bands. 
	The scenario is simpler
		for $\Delta=0$, where particle and hole equations separate; in this case $n_+$ and $n_-$ correspond to different bands, and, just by filling the respective Fermi seas up to the chemical potential, one obtains the two Fermi momentum 
		$k_{F\pm}=\sqrt{k_\mu^2+2 k_\ell ^2\pm\sqrt{4\,k_\ell ^2\,(k_\ell ^2+k_\mu^2)+k_\nu^4}}$, where $k_\nu=\sqrt{2m\nu}/\hbar$. In the absence of SO coupling and the Zeeman field, i.e., $k_\ell=\nu=0$, $k_{\rm F +}=k_{\rm F -}\equiv k_\mu=\sqrt{2m\mu}/\hbar$ is the usual Fermi momentum.  
		When the interparticle interactions operate ($\Delta\neq 0$), the Fermi wave vectors evolve into the minima of the dispersion curves \cite{GiorginiRMP2008}. 
		In particular, in the presence of SO coupling, they can be obtained, with $k_{F_{+}}\ge k_{F_{-}}$, from the lowest positive-energy band of the interacting system as $\partial E_{1}(k) /\partial k=0$ \cite{footnoteFS}.
		Notably, for $\nu \geq \nu_c$, it gives $k_{F-}=0$.
	
	The existence of two Fermi surfaces can be clearly seen from the momentum distribution $ N_k=v_{\downarrow k}^2 + v_{\uparrow k}^2$ for the two bands $E_{1,2}(k)$ of the interacting system, along with the associated (symmetric) condensation amplitude $F_k= u_{\downarrow k}v_{\uparrow k}^*-u_{\uparrow k}v_{\downarrow k}^*$~\cite{deGennes} (Fig.\ref{fig:FGso}). The momentum distribution presents a balanced spin population, since $u_{ \sigma k} = u_{\bar\sigma, \,-k}$ and $v_{ \sigma k} = -v_{\bar\sigma, \,-k}$.
	Before the topological transition, $\nu<\nu_c$,  the lowest energy band $E_1(k)$ gives rise to two separated, inner (contributing to $n_-$) and outer (contributing to $n_+$), regions of occupied momentum states (top panel of Fig.~\ref{fig:FGso}). The highest energy band $E_2(k)$ presents a single momentum region (contributing to $n_+$) of occupied states in the range of wave numbers $k\in[-k_\mu,k_\mu]$. Correspondingly, the  peaks of the condensation amplitude $F_{k}$ appear at $\pm k_{F\pm}$ (top panel of Fig.~\ref{fig:FGso}).
	When the system enters the topological regime ($\nu \ge \nu_c $), the inner momentum region of occupied states in $E_1(k)$ vanishes, and so does the associated condensation amplitude peak occurring  at $k_{F_{-}}=0$ (bottom panel of Fig.~\ref{fig:FGso}).

	\begin{figure}[t]
		\flushleft \textbf{(a)}\\
		\includegraphics[width=1.0\linewidth]{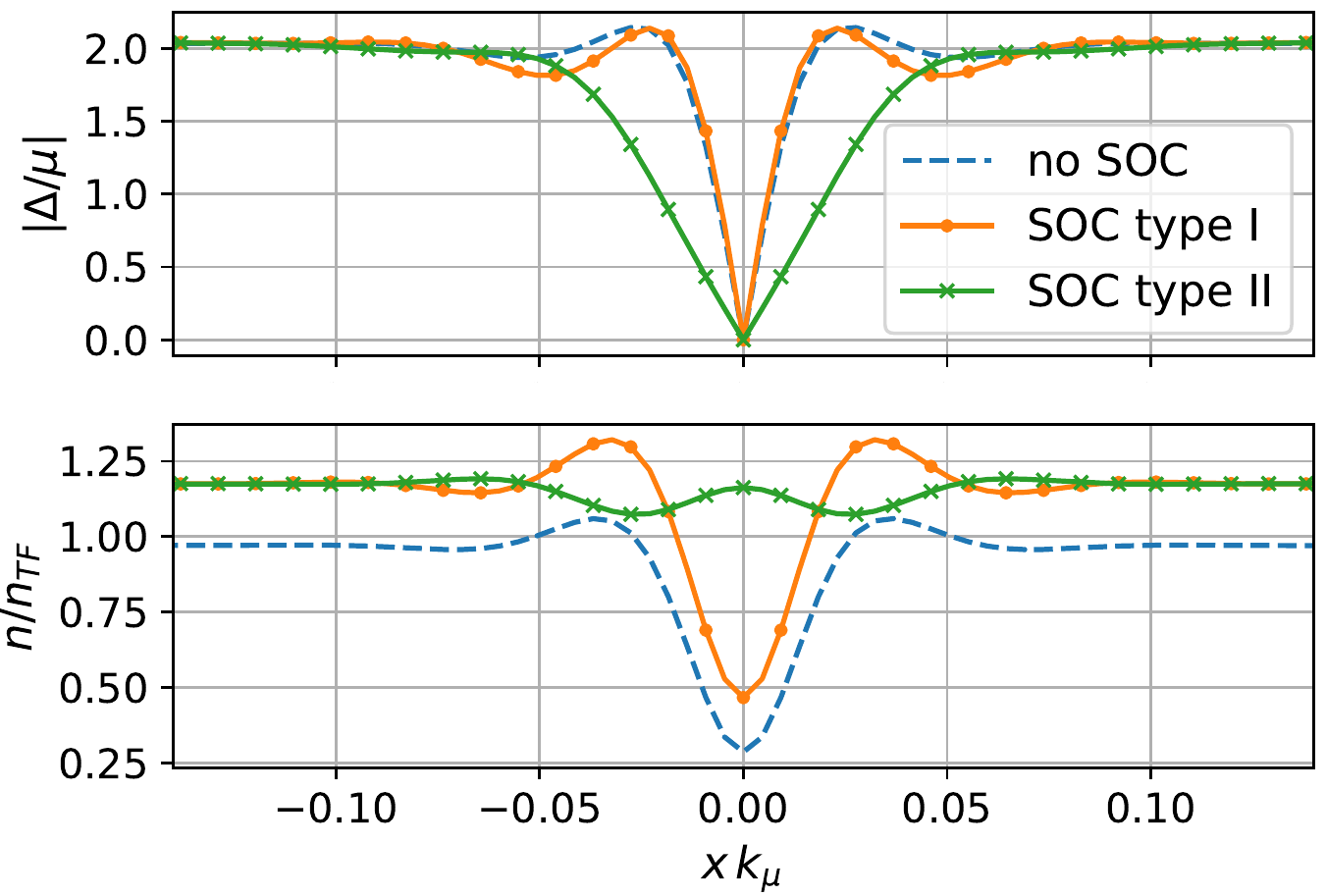}
		\vspace*{-0.7cm}
		\flushleft	\textbf{(b)}\\
		\includegraphics[width=1.0\linewidth]{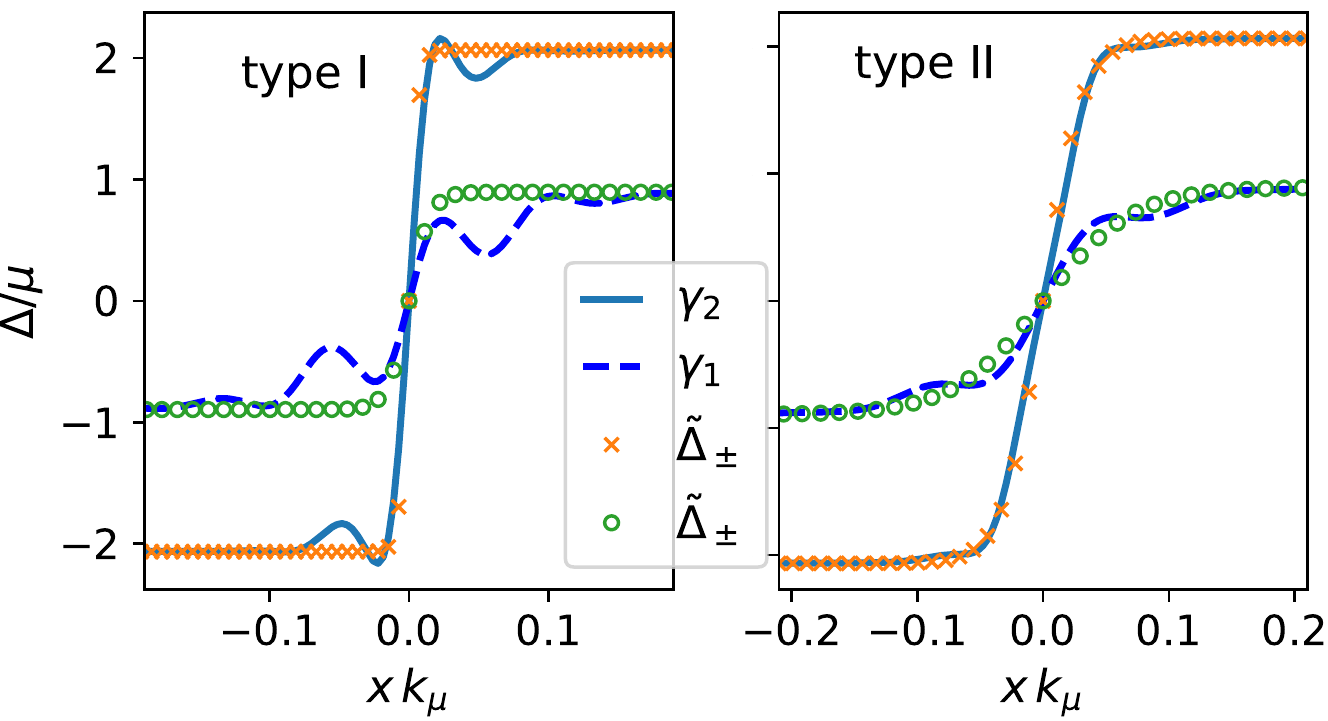}
		\caption{Comparison between Type-I and Type-II solitons in the normal superfluid phase for $\nu=0.53\mu$ and $k_\ell=0.75 k_\mu$. (a) Modulus of the order parameter (top panel) and the density (bottom panel) profiles of type-I and type-II solitons for an interaction strength $\gamma=0.73$. A regular soliton for the same  interaction strength and without SO coupling ($k_\ell=\nu=0$) is also shown for comparison. (b) Order parameter profiles for two interaction values $\gamma_1=0.5$ and $\gamma_2=0.73$. Apart from the Friedel oscillations, the ansatzes $\tilde{\Delta}_{\pm}$ (see the main text) capture well the soliton length scales as probed by the Fermi wave numbers $k_{F-}$ (left panel) and $k_{F+}$ (right panel).  }
		\label{fig:DStypes}
	\end{figure}

\begin{figure}[t]
	\flushleft 
	\includegraphics[width=1.0\linewidth]{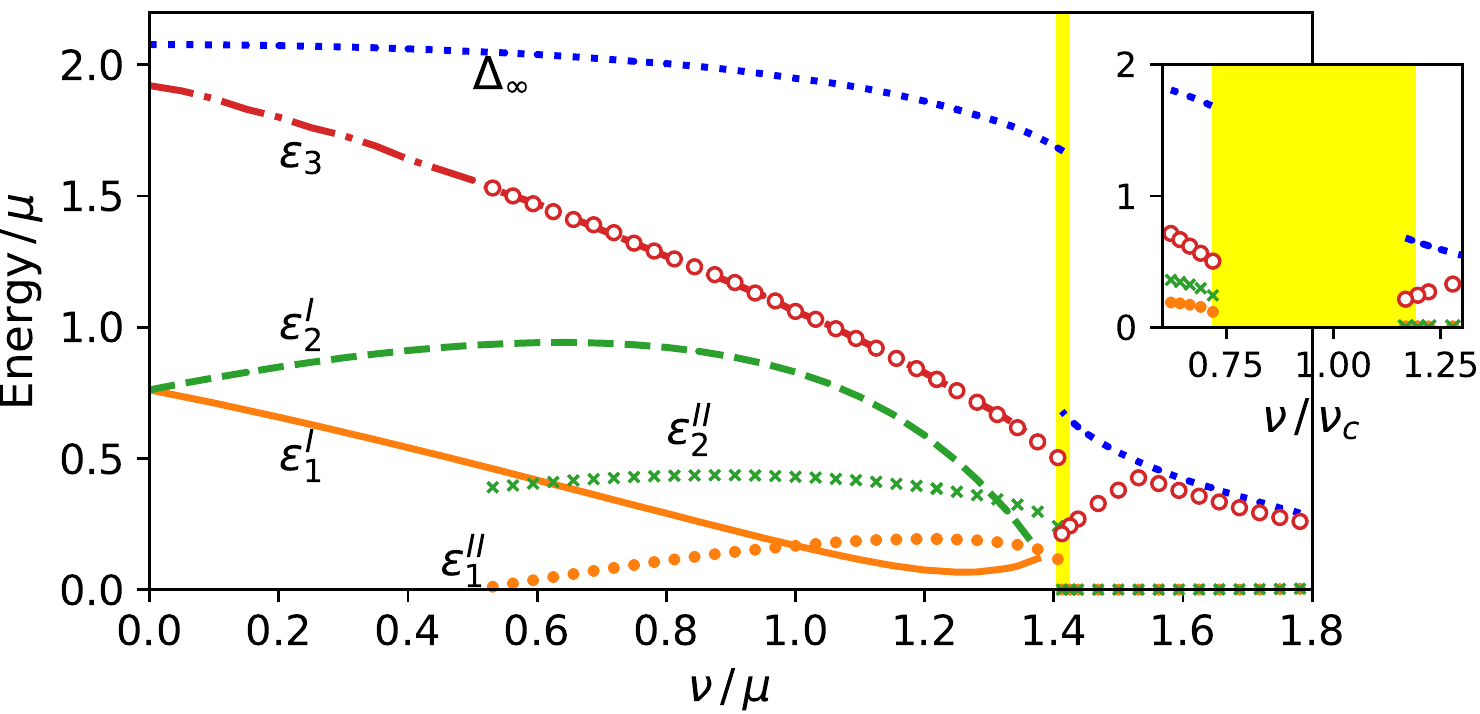}
	\caption{The three lowest quasiparticle energies of the hard-wall trapped system with a dark soliton in the center. Both the interaction strength $\gamma=0.73$ and the SO wavevector $k_\ell=0.75\,k_\mu$ are fixed for varying linear coupling $\nu$.  The topological transition takes place at $\nu_c\approx 1.45\mu$. Below $\nu_c$ the two lowest energy modes are ABSs localized at the soliton core:  $\epsilon^{\rm I}_{1,2}$ and $\epsilon^{\rm II}_{1,2}$ are the energies associated with type-I and type-II solitons, respectively. $\epsilon_3$ is the third lowest excitation energy that corresponds to a bulk mode.  For these parameters, type II solitons emerge at $\nu=\nu_{*} \simeq 0.5\mu$. The inset shows the spectrum in the vicinity of the transition point, where the $x$-axis represents $\nu/\nu_c$ instead of $\nu/\mu$.  In the yellow region the system is  very sensitive to small $\nu$ variations, and the numerical solutions (not shown) present a poor convergence.}
	\label{fig:spectrum}
\end{figure}

	These features suggest that the considered SO-coupled Fermi gas system can support two type of solitons with typical length scales associated with the two values of the Fermi momentum $\xi_\pm=\hbar^2 k_{F\pm}/(m|\Delta_\infty|)$~\cite{Antezza2007}.  We refer to these solutions, associated with $k_{F-}$ and $k_{F+}$, as type-I and type-II solitons, respectively.  The type-I soliton  smoothly connects to the normal dark soliton as $k_\ell\rightarrow 0$ and $\nu \rightarrow 0$. 
	To show that this is the case, we numerically solve the BdG Eqs. (\ref{eq:BdG}) for a system in a hard-wall potential~\cite{footnoteHW}, and search for a self-consistent solution (by means of a modified
		Broyden's method \cite{Johnson1988})
	 starting from the ansatz $\tilde \Delta_-=|\Delta_\infty|\,\tanh(2x/\xi_-)$. We find that the profiles of the order parameter $\Delta$ and the density $n$ of the type-I soliton have similar shapes as those of solitons in the absence of SO coupling for equal interaction, and the small differences are merely quantitative [Fig.~\ref{fig:DStypes} (a)].  
	Our results for the ABSs energies of type I solitons, as functions of $\nu$, are consistent with previous studies~\cite{Liu2015,Xu2014}.  
		Slightly before the topological transition,  the first two ABSs energies become again degenerate (the degeneracy happens also at $\nu=0$) (Fig.~\ref{fig:spectrum}). Beyond this point, we have not found  type-I soliton solutions, which is consistent with the fact of the vanishing Fermi surface associated with $k_{F-}$.   

	\textit{Type-II dark solitons.---}
	We find type-II soliton solutions to the BdG Eqs.~\eqref{eq:BdG} by starting the usual self-consistent numerical procedure from the ansatz $\tilde \Delta_+=|\Delta_\infty|\,\tanh(2x/\xi_+)$. As can be seen in Fig.~\ref{fig:DStypes} (a), not only the widths of the two types of solitons are distinct, but also the presence of Friedel oscillations, notably accentuated in the type-I soliton, marks an important difference between them [Fig.~\ref{fig:DStypes} (b)]. Moreover, the density dip at the core shows a stark contrast between the solitons, with the type-II density having a very low depletion due to the soliton presence [Fig.~\ref{fig:DStypes} (a)].
	Since the length scale $k_{\rm F+}$ persists across the topological transition, the
	associated type-II solitons can be found in both the non-topological and the topological regimes, and so it gives rise to topological solitons that support MZMs.

	In the non-topological regime both types of solitons host two ABSs localized at their cores~\cite{footenoteABS}, whose energies are the lowest among the quasiparticles excitation energies (Fig.~\ref{fig:spectrum}).  
	The lowest energy bound state of type I swaps the $u_{\uparrow j}$ and $v_{\downarrow j}$ components of type II, while the second lowest bound state has essentially the same profile for both types (although their energies differ due to the respective order parameters at the core).
	The other spin components, $u_{\downarrow j}$ and $v_{\uparrow j}$ show equal modulus  $|u_{\uparrow j}(x)|=|u_{\downarrow j}(x)|$ and $|v_{\uparrow j}(x)|=|v_{\downarrow j}(x)|$, but opposite phase gradient 
	$\partial_x\arg u_{\uparrow j}=-\partial_x \arg u_{\downarrow j}$, and
	$\partial_x\arg v_{\uparrow j}=-\partial_x \arg v_{\downarrow j}$.

	Our numerical results show that, for general parameters and given $k_\ell$, there is a threshold linear coupling $\nu_*$, such that type II solitons only exist for $\nu >\nu_*$; below this threshold only type I solitons can be found. In low-pairing systems, $|\Delta|/\mu\ll 1$, the presence of two well-resolved condensation peaks at the Fermi surfaces, i.e. $(k_{F+}-k_{F-}) \xi_+ \gg 1 $ \cite{footnoteII}, is a sufficient condition for the existence of type-II solitons.

	\begin{figure}[t]
		\flushleft \textbf{(a)}
		\includegraphics[width=1.0\linewidth]{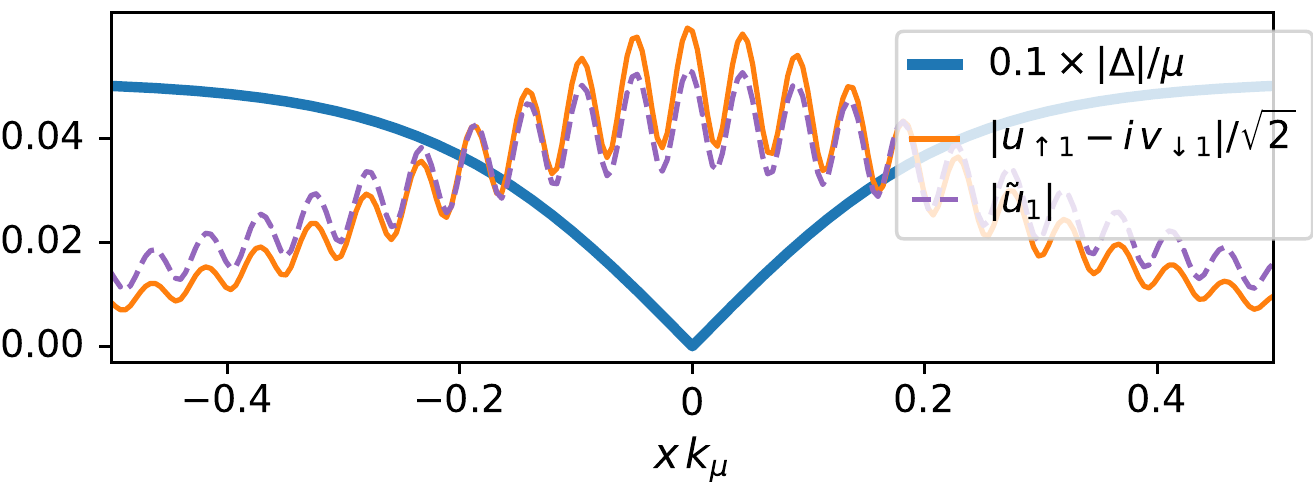}
		\vspace*{-0.8cm}
		\flushleft \textbf{(b)}
		\includegraphics[width=1.015\linewidth]{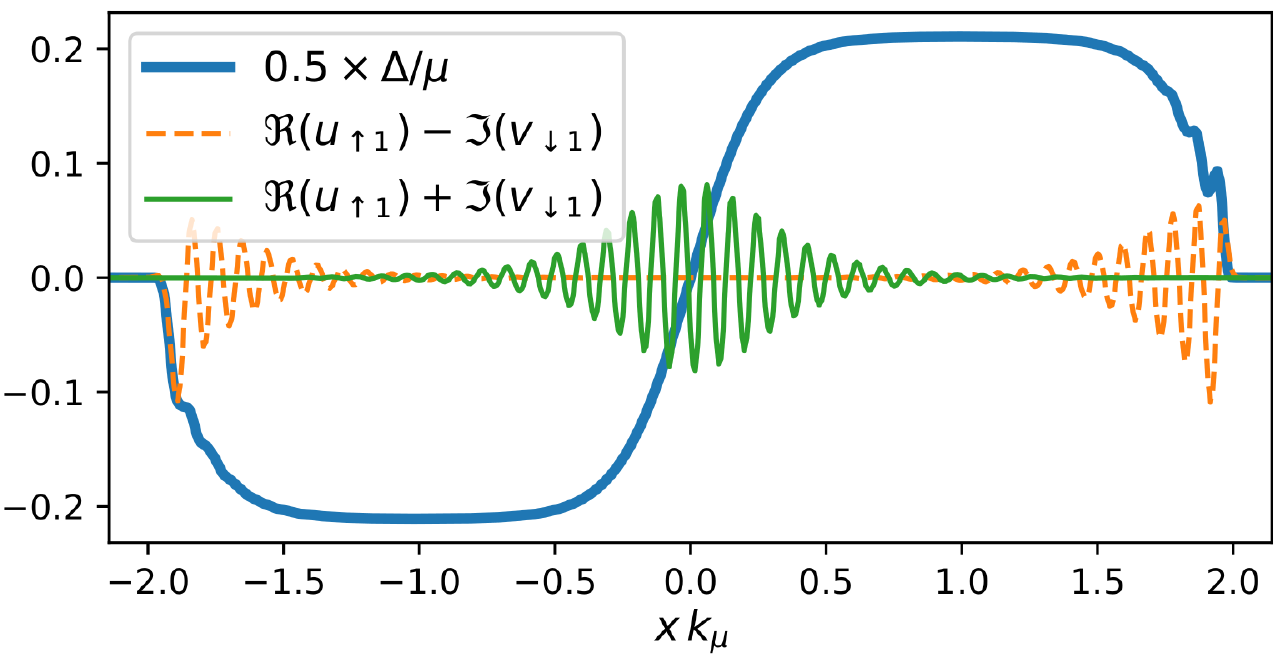}	
		\caption{Type II solitons and MZMs in the topological regime of a hard-wall trapped Fermi gas with interaction strength $\gamma=0.73$, SOC wavevector $k_\ell=0.75\,k_\mu$, and linear coupling $\nu=1.5\,\mu$. (a) Comparison between the numerical solution for one of the MZMs at the soliton core and the analytical ansatz Eq. (\ref{eq:MZMansatz}), evaluated with the value of $k_{F+}$ for the non-interacting gas. (b) The zero-energy modes give rise to localized MZMs at the edges and at the soliton core. Inside the core the modes fulfill $u_{\uparrow}=-i v_{\downarrow}$ while at the edges $u_{\uparrow}=i v_{\downarrow}$ ~\cite{footnoteMZM}. }
		\label{fig:DStypeII}
	\end{figure}

	\textit{Non-topological quasi-zero modes and MZMs.---} Interestingly, at the point of emergence of the type-II soliton $\nu=\nu_{*}$, the lowest energy bound state presents a vanishing excitation energy  (typically  $\epsilon^{\rm II}_{1}(\nu_*) = 10^{-2}\text{---}10^{-3}\, \mu$, see Fig.~\ref{fig:spectrum}). However, this quasi-zero-energy mode does not share the particle-hole symmetry of MZMs, i.e., $|u_{\uparrow j}|\neq |v_{\uparrow j}^*|$. For a zero-energy eigenmode, the particle-hole symmetry relates  $\psi_{0+}$ and $\psi_{0-}$  
	through	$C \psi^{*}_{0+} = \psi_{0-}$.  However for non-topological modes $\psi_{-0}\neq \psi_{+0}$, while for MZMs, within numerical accuracy, $\psi_{0-}\simeq\psi_{0+}$.  The vanishing energy of $\epsilon^{\rm II}_{1}(\nu_*)$ is a result of the accidental cancellation of different energy terms in the BdG hamiltonian and has nothing to do with MZMs (for which we find typical excitation energies several order of magnitude smaller than $\epsilon^{\rm II}_{1}(\nu_*)$). 
	This kind of non-topological quasi-zero mode has also been discovered in other relevant systems~\cite{Brouwer2012,Jorge2015,Liu2017,Fleckenstein2018,Moore2018, Christopher2018,Woods2019,Liu2019,Chen2019,Christian2020}. 
	When $\nu \rightarrow \nu_c$, the two ABSs energies associated with the type-II soliton $\epsilon^{\rm II}_{1,2} \rightarrow 0$ (Fig.~\ref{fig:spectrum}).
	
	\textit{Topological regime.---}
	Within the topological regime, as in previous works~\cite{Xu2014,Liu2015}, we find two fermionic zero-energy eigenstates of the BdG Eqs~\eqref{eq:BdG} with energies $E^{0}_1\sim E^{0}_2 \approx 0$.  Each of the eigenstates can be  decomposed into two MZMs. At the soliton core there are two localized MZMs and the other two  MZMs are localized at the left and right edges (Fig.~\ref{fig:DStypeII}).   
	The MZMs at the soliton core can be written as
	$\psi_{1,2}^M=\mathcal{N}_0\,[U_0,V_0]^T$, where $U_0=[ u_{1,2}, \,  u_{1,2}^*]^T$, $V_0= i\,\sigma_x U_0$ and $\mathcal{N}_0$ is the normalization factor. Here  $u_{1,2}\equiv u_{\uparrow,1,2}$ and we have used the spin balance condition ($u_{\downarrow}=u^{*}_{\uparrow}$,  $v_{\downarrow}=-v^{*}_{\uparrow}$) and the reality condition ($u_{\sigma}=e^{i \phi} v^{*}_{\sigma}$ with $\phi=\pi/2$).
	We propose the following ansatzes of the MZMs at the soliton core: 
	\begin{equation}
	\begin{aligned}
	\tilde u_{1}&={\cal N}_1\, {f(k_{F+}x)}\,\mathrm{sech}(x/\xi_+), \\
	\tilde u_{2}&= -{\cal N}_2\, k_{F+}^{-1}\partial_x f(k_{F+}x) \,\mathrm{sech}(x/\xi_+),
	\label{eq:MZMansatz}
	\end{aligned}
	\end{equation}
	where $f(k_{F+}x)=\cos(k_{F+}x)+i\,\alpha\,\sin(k_{F+}x)$, and  it solves Eq.~\eqref{eq:BdG} exactly with $\alpha=-2k_\ell k_{F+}/(k_\nu^2+k_{F+}^2-k_\mu^2)$ when $\Delta=0$. ${\cal N}_{1,2}$ are normalization constants that produce $\int \,dx \, |\tilde u_{1,2}|^2=1/4$.
	The ansatzes show good agreement with the numerical results [Fig.~\ref{fig:DStypeII}(b)]. In general, two MZMs have to be far apart to avoid the overlapping of their wavefunctions, hence to ensure that their splitting energy is exponentially small. Here the two MZMs are localized in the same core region, but the out of phase oscillation of their wavefunctions produces a vanishing overlap, i.e., $\int\,dx \, (\psi^{M}_1)^{\dagger} \psi^{M}_2=4 \Re (\int dx \, u^*_{1} u_{2})=0$. This phenomenon has also been reported in Refs.~\cite{Xu2014,Fan2019}.

\textit{Conclusion.---}
We discovered a novel type (type-II) of dark solitons in a spin-orbit coupled Fermi gas under an external Zeeman field. Type-II solitons have no correspondence in ordinary Fermi gases and appear only for a finite Zeeman field.  
Previously, the Majorana solitons had been presented in the literature
as the natural counterpart of the regular type-I soliton found in the non-topological
regime. We show that this is not the case, since only the novel type-II soliton exists in both the non-topological and the topological regimes, and so in the latter regime it hosts Majorana zero modes.  
Our findings provide a new scenario of soliton excitations in spin-orbit coupled Fermi gases.  
More generally, the emergence of the type-II soliton in the non-topological regime implies the coexistence, for a given set of parameters in an interacting, quantum-degenerate fermionic system, of two different types of nonlinear excitations featuring a localized
	$\pi$-phase jump in the order parameter. In this regard, type-II solitons
	could also be found in other condensed matter systems in the search for
	the realization of Majorana zero modes, such as the 1D
	hybrid nanowires with a semiconductor-superconductor structure in the presence of
	spin-orbit coupling, where a $\pi$ Josephson junction gives rise to a domain wall
	in the order parameter~\cite{Sau2021}.

The two types of solitons are expected to exhibit strikingly distinct dynamical behaviors.
In contrast to the type-I soliton, the physical mass of the type-II soliton, which accounts for the density dip, is negligible.  For instance, in a harmonic trap, a type-I soliton would oscillate around the potential minimum with a frequency that is still governed by the ratio of its inertial and physical masses~\cite{Scott2011}. While for type-II soliton, spin-orbit and coherent couplings would dominate the motion. Moreover, different soliton generation strategies~\cite{Kong2021} could be required for their experimental realization. Simultaneous amplitude and phase engineering method~\cite{Fritsch2020} might provide an initial density profile more consistent with each soliton type. Detection and identification of the two types of solitons in ultracold-gas 
experiments requires 
probing both the fermionic density and the order parameter. The density could be 
reconstructed via, for instance, phase constrast imaging, while the order parameter 
could be determined by quasiparticle spectroscopy \cite{Schirotzek2008}. 
From these mesurements,
the typical length scales and density depletion of the solitons can be extracted.
An indirect detection of the Majorana modes would be associated 
with the reconstruction of the hosting soliton once the system has entered the 
topological regime. A direct (static) detection of Majorana zero modes would involve 
the resolution of the density of states, or at least, as happens with the weak link 
conductance in hybrid nanowires \cite{Sau2021}, the measurement of a transport quantity
capable of probing the density of states. In this regard, an anomalous result from the mesurement of the current-phase Josephson current across the soliton could provide the signature of the presence of Majorana zero modes~\cite{Asano2006}.

\textit{Acknowledgments---}
We thank J. Brand and P. Brydon for their careful reading of early versions of this manuscript and their valuable feedback. We thanks P. Zou and P.B. Blakie for useful discussions.  X.Y. acknowledges the support from NSAF (No. U1930403) and NSFC (No. 12175215). A.M.M gratefully acknowledges the hospitality of Graduate School of China Academy of Engineering Physics, while part of this work was accomplished.


\begin{thebibliography}{60}%
	\makeatletter
	\providecommand \@ifxundefined [1]{%
		\@ifx{#1\undefined}
	}%
	\providecommand \@ifnum [1]{%
		\ifnum #1\expandafter \@firstoftwo
		\else \expandafter \@secondoftwo
		\fi
	}%
	\providecommand \@ifx [1]{%
		\ifx #1\expandafter \@firstoftwo
		\else \expandafter \@secondoftwo
		\fi
	}%
	\providecommand \natexlab [1]{#1}%
	\providecommand \enquote  [1]{``#1''}%
	\providecommand \bibnamefont  [1]{#1}%
	\providecommand \bibfnamefont [1]{#1}%
	\providecommand \citenamefont [1]{#1}%
	\providecommand \href@noop [0]{\@secondoftwo}%
	\providecommand \href [0]{\begingroup \@sanitize@url \@href}%
	\providecommand \@href[1]{\@@startlink{#1}\@@href}%
	\providecommand \@@href[1]{\endgroup#1\@@endlink}%
	\providecommand \@sanitize@url [0]{\catcode `\\12\catcode `\$12\catcode
		`\&12\catcode `\#12\catcode `\^12\catcode `\_12\catcode `\%12\relax}%
	\providecommand \@@startlink[1]{}%
	\providecommand \@@endlink[0]{}%
	\providecommand \url  [0]{\begingroup\@sanitize@url \@url }%
	\providecommand \@url [1]{\endgroup\@href {#1}{\urlprefix }}%
	\providecommand \urlprefix  [0]{URL }%
	\providecommand \Eprint [0]{\href }%
	\providecommand \doibase [0]{http://dx.doi.org/}%
	\providecommand \selectlanguage [0]{\@gobble}%
	\providecommand \bibinfo  [0]{\@secondoftwo}%
	\providecommand \bibfield  [0]{\@secondoftwo}%
	\providecommand \translation [1]{[#1]}%
	\providecommand \BibitemOpen [0]{}%
	\providecommand \bibitemStop [0]{}%
	\providecommand \bibitemNoStop [0]{.\EOS\space}%
	\providecommand \EOS [0]{\spacefactor3000\relax}%
	\providecommand \BibitemShut  [1]{\csname bibitem#1\endcsname}%
	\let\auto@bib@innerbib\@empty
	\bibitem [{\citenamefont {Frantzeskakis}(2010)}]{Frantzeskakis2010}%
	\BibitemOpen
	\bibfield  {author} {\bibinfo {author} {\bibfnamefont {D.~J.}\ \bibnamefont
			{Frantzeskakis}},\ }\href {\doibase 10.1088/1751-8113/43/21/213001}
	{\bibfield  {journal} {\bibinfo  {journal} {Journal of Physics A:
				Mathematical and Theoretical}\ }\textbf {\bibinfo {volume} {43}},\ \bibinfo
		{pages} {213001} (\bibinfo {year} {2010})}\BibitemShut {NoStop}%
	\bibitem [{\citenamefont {Giorgini}\ \emph {et~al.}(2008)\citenamefont
		{Giorgini}, \citenamefont {Pitaevskii},\ and\ \citenamefont
		{Stringari}}]{GiorginiRMP2008}%
	\BibitemOpen
	\bibfield  {author} {\bibinfo {author} {\bibfnamefont {S.}~\bibnamefont
			{Giorgini}}, \bibinfo {author} {\bibfnamefont {L.~P.}\ \bibnamefont
			{Pitaevskii}}, \ and\ \bibinfo {author} {\bibfnamefont {S.}~\bibnamefont
			{Stringari}},\ }\href {\doibase 10.1103/RevModPhys.80.1215} {\bibfield
		{journal} {\bibinfo  {journal} {Rev. Mod. Phys.}\ }\textbf {\bibinfo {volume}
			{80}},\ \bibinfo {pages} {1215} (\bibinfo {year} {2008})}\BibitemShut
	{NoStop}%
	\bibitem [{\citenamefont {Antezza}\ \emph {et~al.}(2007)\citenamefont
		{Antezza}, \citenamefont {Dalfovo}, \citenamefont {Pitaevskii},\ and\
		\citenamefont {Stringari}}]{Antezza2007}%
	\BibitemOpen
	\bibfield  {author} {\bibinfo {author} {\bibfnamefont {M.}~\bibnamefont
			{Antezza}}, \bibinfo {author} {\bibfnamefont {F.}~\bibnamefont {Dalfovo}},
		\bibinfo {author} {\bibfnamefont {L.~P.}\ \bibnamefont {Pitaevskii}}, \ and\
		\bibinfo {author} {\bibfnamefont {S.}~\bibnamefont {Stringari}},\ }\href
	{\doibase 10.1103/PhysRevA.76.043610} {\bibfield  {journal} {\bibinfo
			{journal} {Phys. Rev. A}\ }\textbf {\bibinfo {volume} {76}},\ \bibinfo
		{pages} {043610} (\bibinfo {year} {2007})}\BibitemShut {NoStop}%
	\bibitem [{\citenamefont {Scott}\ \emph {et~al.}(2011)\citenamefont {Scott},
		\citenamefont {Dalfovo}, \citenamefont {Pitaevskii},\ and\ \citenamefont
		{Stringari}}]{Scott2011}%
	\BibitemOpen
	\bibfield  {author} {\bibinfo {author} {\bibfnamefont {R.~G.}\ \bibnamefont
			{Scott}}, \bibinfo {author} {\bibfnamefont {F.}~\bibnamefont {Dalfovo}},
		\bibinfo {author} {\bibfnamefont {L.~P.}\ \bibnamefont {Pitaevskii}}, \ and\
		\bibinfo {author} {\bibfnamefont {S.}~\bibnamefont {Stringari}},\ }\href
	{\doibase 10.1103/PhysRevLett.106.185301} {\bibfield  {journal} {\bibinfo
			{journal} {Phys. Rev. Lett.}\ }\textbf {\bibinfo {volume} {106}},\ \bibinfo
		{pages} {185301} (\bibinfo {year} {2011})}\BibitemShut {NoStop}%
	\bibitem [{\citenamefont {Liao}\ and\ \citenamefont {Brand}(2011)}]{Liao2011}%
	\BibitemOpen
	\bibfield  {author} {\bibinfo {author} {\bibfnamefont {R.}~\bibnamefont
			{Liao}}\ and\ \bibinfo {author} {\bibfnamefont {J.}~\bibnamefont {Brand}},\
	}\href {\doibase 10.1103/PhysRevA.83.041604} {\bibfield  {journal} {\bibinfo
			{journal} {Phys. Rev. A}\ }\textbf {\bibinfo {volume} {83}},\ \bibinfo
		{pages} {041604} (\bibinfo {year} {2011})}\BibitemShut {NoStop}%
	\bibitem [{\citenamefont {Spuntarelli}\ \emph {et~al.}(2011)\citenamefont
		{Spuntarelli}, \citenamefont {Carr}, \citenamefont {Pieri},\ and\
		\citenamefont {Strinati}}]{Spuntarelli2011}%
	\BibitemOpen
	\bibfield  {author} {\bibinfo {author} {\bibfnamefont {A.}~\bibnamefont
			{Spuntarelli}}, \bibinfo {author} {\bibfnamefont {L.~D.}\ \bibnamefont
			{Carr}}, \bibinfo {author} {\bibfnamefont {P.}~\bibnamefont {Pieri}}, \ and\
		\bibinfo {author} {\bibfnamefont {G.~C.}\ \bibnamefont {Strinati}},\ }\href
	{\doibase 10.1088/1367-2630/13/3/035010} {\bibfield  {journal} {\bibinfo
			{journal} {New Journal of Physics}\ }\textbf {\bibinfo {volume} {13}},\
		\bibinfo {pages} {035010} (\bibinfo {year} {2011})}\BibitemShut {NoStop}%
	\bibitem [{\citenamefont {Cetoli}\ \emph {et~al.}(2013)\citenamefont {Cetoli},
		\citenamefont {Brand}, \citenamefont {Scott}, \citenamefont {Dalfovo},\ and\
		\citenamefont {Pitaevskii}}]{Cetoli2013}%
	\BibitemOpen
	\bibfield  {author} {\bibinfo {author} {\bibfnamefont {A.}~\bibnamefont
			{Cetoli}}, \bibinfo {author} {\bibfnamefont {J.}~\bibnamefont {Brand}},
		\bibinfo {author} {\bibfnamefont {R.~G.}\ \bibnamefont {Scott}}, \bibinfo
		{author} {\bibfnamefont {F.}~\bibnamefont {Dalfovo}}, \ and\ \bibinfo
		{author} {\bibfnamefont {L.~P.}\ \bibnamefont {Pitaevskii}},\ }\href
	{\doibase 10.1103/PhysRevA.88.043639} {\bibfield  {journal} {\bibinfo
			{journal} {Phys. Rev. A}\ }\textbf {\bibinfo {volume} {88}},\ \bibinfo
		{pages} {043639} (\bibinfo {year} {2013})}\BibitemShut {NoStop}%
	\bibitem [{\citenamefont {Efimkin}\ and\ \citenamefont
		{Galitski}(2015)}]{Efimkin2015}%
	\BibitemOpen
	\bibfield  {author} {\bibinfo {author} {\bibfnamefont {D.~K.}\ \bibnamefont
			{Efimkin}}\ and\ \bibinfo {author} {\bibfnamefont {V.}~\bibnamefont
			{Galitski}},\ }\href {\doibase 10.1103/PhysRevA.91.023616} {\bibfield
		{journal} {\bibinfo  {journal} {Phys. Rev. A}\ }\textbf {\bibinfo {volume}
			{91}},\ \bibinfo {pages} {023616} (\bibinfo {year} {2015})}\BibitemShut
	{NoStop}%
	\bibitem [{\citenamefont {Yefsah}\ \emph {et~al.}(2013)\citenamefont {Yefsah},
		\citenamefont {Sommer}, \citenamefont {Ku}, \citenamefont {Cheuk},
		\citenamefont {Ji}, \citenamefont {Bakr},\ and\ \citenamefont
		{Zwierlein}}]{Yefsah2013}%
	\BibitemOpen
	\bibfield  {author} {\bibinfo {author} {\bibfnamefont {T.}~\bibnamefont
			{Yefsah}}, \bibinfo {author} {\bibfnamefont {A.~T.}\ \bibnamefont {Sommer}},
		\bibinfo {author} {\bibfnamefont {M.~J.}\ \bibnamefont {Ku}}, \bibinfo
		{author} {\bibfnamefont {L.~W.}\ \bibnamefont {Cheuk}}, \bibinfo {author}
		{\bibfnamefont {W.}~\bibnamefont {Ji}}, \bibinfo {author} {\bibfnamefont
			{W.~S.}\ \bibnamefont {Bakr}}, \ and\ \bibinfo {author} {\bibfnamefont
			{M.~W.}\ \bibnamefont {Zwierlein}},\ }\href {\doibase 10.1038/nature12338}
	{\bibfield  {journal} {\bibinfo  {journal} {Nature}\ }\textbf {\bibinfo
			{volume} {499}},\ \bibinfo {pages} {426} (\bibinfo {year}
		{2013})}\BibitemShut {NoStop}%
	\bibitem [{\citenamefont {Ku}\ \emph {et~al.}(2016)\citenamefont {Ku},
		\citenamefont {Mukherjee}, \citenamefont {Yefsah},\ and\ \citenamefont
		{Zwierlein}}]{Ku2014}%
	\BibitemOpen
	\bibfield  {author} {\bibinfo {author} {\bibfnamefont {M.~J.~H.}\
			\bibnamefont {Ku}}, \bibinfo {author} {\bibfnamefont {B.}~\bibnamefont
			{Mukherjee}}, \bibinfo {author} {\bibfnamefont {T.}~\bibnamefont {Yefsah}}, \
		and\ \bibinfo {author} {\bibfnamefont {M.~W.}\ \bibnamefont {Zwierlein}},\
	}\href {\doibase 10.1103/PhysRevLett.116.045304} {\bibfield  {journal}
		{\bibinfo  {journal} {Phys. Rev. Lett.}\ }\textbf {\bibinfo {volume} {116}},\
		\bibinfo {pages} {045304} (\bibinfo {year} {2016})}\BibitemShut {NoStop}%
	\bibitem [{\citenamefont {Ku}\ \emph {et~al.}(2014)\citenamefont {Ku},
		\citenamefont {Ji}, \citenamefont {Mukherjee}, \citenamefont
		{Guardado-Sanchez}, \citenamefont {Cheuk}, \citenamefont {Yefsah},\ and\
		\citenamefont {Zwierlein}}]{Zwierleinprl2014}%
	\BibitemOpen
	\bibfield  {author} {\bibinfo {author} {\bibfnamefont {M.~J.~H.}\
			\bibnamefont {Ku}}, \bibinfo {author} {\bibfnamefont {W.}~\bibnamefont {Ji}},
		\bibinfo {author} {\bibfnamefont {B.}~\bibnamefont {Mukherjee}}, \bibinfo
		{author} {\bibfnamefont {E.}~\bibnamefont {Guardado-Sanchez}}, \bibinfo
		{author} {\bibfnamefont {L.~W.}\ \bibnamefont {Cheuk}}, \bibinfo {author}
		{\bibfnamefont {T.}~\bibnamefont {Yefsah}}, \ and\ \bibinfo {author}
		{\bibfnamefont {M.~W.}\ \bibnamefont {Zwierlein}},\ }\href {\doibase
		10.1103/PhysRevLett.113.065301} {\bibfield  {journal} {\bibinfo  {journal}
			{Phys. Rev. Lett.}\ }\textbf {\bibinfo {volume} {113}},\ \bibinfo {pages}
		{065301} (\bibinfo {year} {2014})}\BibitemShut {NoStop}%
	\bibitem [{\citenamefont {Wang}\ \emph {et~al.}(2012)\citenamefont {Wang},
		\citenamefont {Yu}, \citenamefont {Fu}, \citenamefont {Miao}, \citenamefont
		{Huang}, \citenamefont {Chai}, \citenamefont {Zhai},\ and\ \citenamefont
		{Zhang}}]{Wang2012}%
	\BibitemOpen
	\bibfield  {author} {\bibinfo {author} {\bibfnamefont {P.}~\bibnamefont
			{Wang}}, \bibinfo {author} {\bibfnamefont {Z.-Q.}\ \bibnamefont {Yu}},
		\bibinfo {author} {\bibfnamefont {Z.}~\bibnamefont {Fu}}, \bibinfo {author}
		{\bibfnamefont {J.}~\bibnamefont {Miao}}, \bibinfo {author} {\bibfnamefont
			{L.}~\bibnamefont {Huang}}, \bibinfo {author} {\bibfnamefont
			{S.}~\bibnamefont {Chai}}, \bibinfo {author} {\bibfnamefont {H.}~\bibnamefont
			{Zhai}}, \ and\ \bibinfo {author} {\bibfnamefont {J.}~\bibnamefont {Zhang}},\
	}\href {\doibase 10.1103/PhysRevLett.109.095301} {\bibfield  {journal}
		{\bibinfo  {journal} {Phys. Rev. Lett.}\ }\textbf {\bibinfo {volume} {109}},\
		\bibinfo {pages} {095301} (\bibinfo {year} {2012})}\BibitemShut {NoStop}%
	\bibitem [{\citenamefont {Cheuk}\ \emph {et~al.}(2012)\citenamefont {Cheuk},
		\citenamefont {Sommer}, \citenamefont {Hadzibabic}, \citenamefont {Yefsah},
		\citenamefont {Bakr},\ and\ \citenamefont {Zwierlein}}]{Cheuk2012}%
	\BibitemOpen
	\bibfield  {author} {\bibinfo {author} {\bibfnamefont {L.~W.}\ \bibnamefont
			{Cheuk}}, \bibinfo {author} {\bibfnamefont {A.~T.}\ \bibnamefont {Sommer}},
		\bibinfo {author} {\bibfnamefont {Z.}~\bibnamefont {Hadzibabic}}, \bibinfo
		{author} {\bibfnamefont {T.}~\bibnamefont {Yefsah}}, \bibinfo {author}
		{\bibfnamefont {W.~S.}\ \bibnamefont {Bakr}}, \ and\ \bibinfo {author}
		{\bibfnamefont {M.~W.}\ \bibnamefont {Zwierlein}},\ }\href {\doibase
		10.1103/PhysRevLett.109.095302} {\bibfield  {journal} {\bibinfo  {journal}
			{Phys. Rev. Lett.}\ }\textbf {\bibinfo {volume} {109}},\ \bibinfo {pages}
		{095302} (\bibinfo {year} {2012})}\BibitemShut {NoStop}%
	\bibitem [{\citenamefont {Fu}\ \emph {et~al.}(2013)\citenamefont {Fu},
		\citenamefont {Huang}, \citenamefont {Meng}, \citenamefont {Wang},
		\citenamefont {Liu}, \citenamefont {Pu}, \citenamefont {Hu},\ and\
		\citenamefont {Zhang}}]{Fu2013}%
	\BibitemOpen
	\bibfield  {author} {\bibinfo {author} {\bibfnamefont {Z.}~\bibnamefont
			{Fu}}, \bibinfo {author} {\bibfnamefont {L.}~\bibnamefont {Huang}}, \bibinfo
		{author} {\bibfnamefont {Z.}~\bibnamefont {Meng}}, \bibinfo {author}
		{\bibfnamefont {P.}~\bibnamefont {Wang}}, \bibinfo {author} {\bibfnamefont
			{X.-J.}\ \bibnamefont {Liu}}, \bibinfo {author} {\bibfnamefont
			{H.}~\bibnamefont {Pu}}, \bibinfo {author} {\bibfnamefont {H.}~\bibnamefont
			{Hu}}, \ and\ \bibinfo {author} {\bibfnamefont {J.}~\bibnamefont {Zhang}},\
	}\href {\doibase 10.1103/PhysRevA.87.053619} {\bibfield  {journal} {\bibinfo
			{journal} {Phys. Rev. A}\ }\textbf {\bibinfo {volume} {87}},\ \bibinfo
		{pages} {053619} (\bibinfo {year} {2013})}\BibitemShut {NoStop}%
	\bibitem [{\citenamefont {Williams}\ \emph {et~al.}(2013)\citenamefont
		{Williams}, \citenamefont {Beeler}, \citenamefont {LeBlanc}, \citenamefont
		{Jim\'enez-Garc\'{\i}a},\ and\ \citenamefont {Spielman}}]{Williams2013}%
	\BibitemOpen
	\bibfield  {author} {\bibinfo {author} {\bibfnamefont {R.~A.}\ \bibnamefont
			{Williams}}, \bibinfo {author} {\bibfnamefont {M.~C.}\ \bibnamefont
			{Beeler}}, \bibinfo {author} {\bibfnamefont {L.~J.}\ \bibnamefont {LeBlanc}},
		\bibinfo {author} {\bibfnamefont {K.}~\bibnamefont {Jim\'enez-Garc\'{\i}a}},
		\ and\ \bibinfo {author} {\bibfnamefont {I.~B.}\ \bibnamefont {Spielman}},\
	}\href {\doibase 10.1103/PhysRevLett.111.095301} {\bibfield  {journal}
		{\bibinfo  {journal} {Phys. Rev. Lett.}\ }\textbf {\bibinfo {volume} {111}},\
		\bibinfo {pages} {095301} (\bibinfo {year} {2013})}\BibitemShut {NoStop}%
	\bibitem [{\citenamefont {Fu}\ \emph {et~al.}(2014)\citenamefont {Fu},
		\citenamefont {Huang}, \citenamefont {Meng}, \citenamefont {Wang},
		\citenamefont {Zhang}, \citenamefont {Zhang}, \citenamefont {Zhai},
		\citenamefont {Zhang},\ and\ \citenamefont {Zhang}}]{Fu2014}%
	\BibitemOpen
	\bibfield  {author} {\bibinfo {author} {\bibfnamefont {Z.}~\bibnamefont
			{Fu}}, \bibinfo {author} {\bibfnamefont {L.}~\bibnamefont {Huang}}, \bibinfo
		{author} {\bibfnamefont {Z.}~\bibnamefont {Meng}}, \bibinfo {author}
		{\bibfnamefont {P.}~\bibnamefont {Wang}}, \bibinfo {author} {\bibfnamefont
			{L.}~\bibnamefont {Zhang}}, \bibinfo {author} {\bibfnamefont
			{S.}~\bibnamefont {Zhang}}, \bibinfo {author} {\bibfnamefont
			{H.}~\bibnamefont {Zhai}}, \bibinfo {author} {\bibfnamefont {P.}~\bibnamefont
			{Zhang}}, \ and\ \bibinfo {author} {\bibfnamefont {J.}~\bibnamefont
			{Zhang}},\ }\href {\doibase 10.1038/nphys3672} {\bibfield  {journal}
		{\bibinfo  {journal} {Nature physics}\ }\textbf {\bibinfo {volume} {10}},\
		\bibinfo {pages} {110} (\bibinfo {year} {2014})}\BibitemShut {NoStop}%
	\bibitem [{\citenamefont {Burdick}\ \emph {et~al.}(2016)\citenamefont
		{Burdick}, \citenamefont {Tang},\ and\ \citenamefont {Lev}}]{Burdick2016}%
	\BibitemOpen
	\bibfield  {author} {\bibinfo {author} {\bibfnamefont {N.~Q.}\ \bibnamefont
			{Burdick}}, \bibinfo {author} {\bibfnamefont {Y.}~\bibnamefont {Tang}}, \
		and\ \bibinfo {author} {\bibfnamefont {B.~L.}\ \bibnamefont {Lev}},\ }\href
	{\doibase 10.1103/PhysRevX.6.031022} {\bibfield  {journal} {\bibinfo
			{journal} {Phys. Rev. X}\ }\textbf {\bibinfo {volume} {6}},\ \bibinfo {pages}
		{031022} (\bibinfo {year} {2016})}\BibitemShut {NoStop}%
	\bibitem [{\citenamefont {Song}\ \emph {et~al.}(2016)\citenamefont {Song},
		\citenamefont {He}, \citenamefont {Zhang}, \citenamefont {Hajiyev},
		\citenamefont {Huang}, \citenamefont {Liu},\ and\ \citenamefont
		{Jo}}]{Song2016}%
	\BibitemOpen
	\bibfield  {author} {\bibinfo {author} {\bibfnamefont {B.}~\bibnamefont
			{Song}}, \bibinfo {author} {\bibfnamefont {C.}~\bibnamefont {He}}, \bibinfo
		{author} {\bibfnamefont {S.}~\bibnamefont {Zhang}}, \bibinfo {author}
		{\bibfnamefont {E.}~\bibnamefont {Hajiyev}}, \bibinfo {author} {\bibfnamefont
			{W.}~\bibnamefont {Huang}}, \bibinfo {author} {\bibfnamefont {X.-J.}\
			\bibnamefont {Liu}}, \ and\ \bibinfo {author} {\bibfnamefont {G.-B.}\
			\bibnamefont {Jo}},\ }\href {\doibase 10.1103/PhysRevA.94.061604} {\bibfield
		{journal} {\bibinfo  {journal} {Phys. Rev. A}\ }\textbf {\bibinfo {volume}
			{94}},\ \bibinfo {pages} {061604} (\bibinfo {year} {2016})}\BibitemShut
	{NoStop}%
	\bibitem [{\citenamefont {Zhai}(2015)}]{Zhai2015}%
	\BibitemOpen
	\bibfield  {author} {\bibinfo {author} {\bibfnamefont {H.}~\bibnamefont
			{Zhai}},\ }\href {\doibase 10.1088/0034-4885/78/2/026001} {\bibfield
		{journal} {\bibinfo  {journal} {Reports on Progress in Physics}\ }\textbf
		{\bibinfo {volume} {78}},\ \bibinfo {pages} {026001} (\bibinfo {year}
		{2015})}\BibitemShut {NoStop}%
	\bibitem [{\citenamefont {Oreg}\ \emph {et~al.}(2010)\citenamefont {Oreg},
		\citenamefont {Refael},\ and\ \citenamefont {von Oppen}}]{Oreg2010}%
	\BibitemOpen
	\bibfield  {author} {\bibinfo {author} {\bibfnamefont {Y.}~\bibnamefont
			{Oreg}}, \bibinfo {author} {\bibfnamefont {G.}~\bibnamefont {Refael}}, \ and\
		\bibinfo {author} {\bibfnamefont {F.}~\bibnamefont {von Oppen}},\ }\href
	{\doibase 10.1103/PhysRevLett.105.177002} {\bibfield  {journal} {\bibinfo
			{journal} {Phys. Rev. Lett.}\ }\textbf {\bibinfo {volume} {105}},\ \bibinfo
		{pages} {177002} (\bibinfo {year} {2010})}\BibitemShut {NoStop}%
	\bibitem [{\citenamefont {Lutchyn}\ \emph {et~al.}(2010)\citenamefont
		{Lutchyn}, \citenamefont {Sau},\ and\ \citenamefont
		{Das~Sarma}}]{Lutchyn2010}%
	\BibitemOpen
	\bibfield  {author} {\bibinfo {author} {\bibfnamefont {R.~M.}\ \bibnamefont
			{Lutchyn}}, \bibinfo {author} {\bibfnamefont {J.~D.}\ \bibnamefont {Sau}}, \
		and\ \bibinfo {author} {\bibfnamefont {S.}~\bibnamefont {Das~Sarma}},\ }\href
	{\doibase 10.1103/PhysRevLett.105.077001} {\bibfield  {journal} {\bibinfo
			{journal} {Phys. Rev. Lett.}\ }\textbf {\bibinfo {volume} {105}},\ \bibinfo
		{pages} {077001} (\bibinfo {year} {2010})}\BibitemShut {NoStop}%
	\bibitem [{\citenamefont {Wei}\ and\ \citenamefont {Mueller}(2012)}]{Wei2012}%
	\BibitemOpen
	\bibfield  {author} {\bibinfo {author} {\bibfnamefont {R.}~\bibnamefont
			{Wei}}\ and\ \bibinfo {author} {\bibfnamefont {E.~J.}\ \bibnamefont
			{Mueller}},\ }\href {\doibase 10.1103/PhysRevA.86.063604} {\bibfield
		{journal} {\bibinfo  {journal} {Phys. Rev. A}\ }\textbf {\bibinfo {volume}
			{86}},\ \bibinfo {pages} {063604} (\bibinfo {year} {2012})}\BibitemShut
	{NoStop}%
	\bibitem [{\citenamefont {Liu}\ and\ \citenamefont {Hu}(2012)}]{LiuHu2012}%
	\BibitemOpen
	\bibfield  {author} {\bibinfo {author} {\bibfnamefont {X.-J.}\ \bibnamefont
			{Liu}}\ and\ \bibinfo {author} {\bibfnamefont {H.}~\bibnamefont {Hu}},\
	}\href {\doibase 10.1103/PhysRevA.85.033622} {\bibfield  {journal} {\bibinfo
			{journal} {Phys. Rev. A}\ }\textbf {\bibinfo {volume} {85}},\ \bibinfo
		{pages} {033622} (\bibinfo {year} {2012})}\BibitemShut {NoStop}%
	\bibitem [{\citenamefont {Liu}\ \emph {et~al.}(2012)\citenamefont {Liu},
		\citenamefont {Jiang}, \citenamefont {Pu},\ and\ \citenamefont
		{Hu}}]{Liu2012}%
	\BibitemOpen
	\bibfield  {author} {\bibinfo {author} {\bibfnamefont {X.-J.}\ \bibnamefont
			{Liu}}, \bibinfo {author} {\bibfnamefont {L.}~\bibnamefont {Jiang}}, \bibinfo
		{author} {\bibfnamefont {H.}~\bibnamefont {Pu}}, \ and\ \bibinfo {author}
		{\bibfnamefont {H.}~\bibnamefont {Hu}},\ }\href {\doibase
		10.1103/PhysRevA.85.021603} {\bibfield  {journal} {\bibinfo  {journal} {Phys.
				Rev. A}\ }\textbf {\bibinfo {volume} {85}},\ \bibinfo {pages} {021603}
		(\bibinfo {year} {2012})}\BibitemShut {NoStop}%
	\bibitem [{\citenamefont {Kitaev}(2001)}]{Kitaev2001}%
	\BibitemOpen
	\bibfield  {author} {\bibinfo {author} {\bibfnamefont {A.~Y.}\ \bibnamefont
			{Kitaev}},\ }\href {\doibase 10.1070/1063-7869/44/10s/s29} {\bibfield
		{journal} {\bibinfo  {journal} {Physics-Uspekhi}\ }\textbf {\bibinfo {volume}
			{44}},\ \bibinfo {pages} {131} (\bibinfo {year} {2001})}\BibitemShut
	{NoStop}%
	\bibitem [{\citenamefont {Alicea}(2012)}]{Alicea2012}%
	\BibitemOpen
	\bibfield  {author} {\bibinfo {author} {\bibfnamefont {J.}~\bibnamefont
			{Alicea}},\ }\href {\doibase 10.1088/0034-4885/75/7/076501} {\bibfield
		{journal} {\bibinfo  {journal} {Reports on Progress in Physics}\ }\textbf
		{\bibinfo {volume} {75}},\ \bibinfo {pages} {076501} (\bibinfo {year}
		{2012})}\BibitemShut {NoStop}%
	\bibitem [{\citenamefont {Aguado}(2017)}]{aguado2017majorana}%
	\BibitemOpen
	\bibfield  {author} {\bibinfo {author} {\bibfnamefont {R.}~\bibnamefont
			{Aguado}},\ }\href@noop {} {\bibfield  {journal} {\bibinfo  {journal} {La
				Rivista del Nuovo Cimento}\ }\textbf {\bibinfo {volume} {40}},\ \bibinfo
		{pages} {523} (\bibinfo {year} {2017})}\BibitemShut {NoStop}%
	\bibitem [{\citenamefont {Kitaev}(2003)}]{kitaev2003fault}%
	\BibitemOpen
	\bibfield  {author} {\bibinfo {author} {\bibfnamefont {A.~Y.}\ \bibnamefont
			{Kitaev}},\ }\href {\doibase 10.1016/S0003-4916(02)00018-0} {\bibfield
		{journal} {\bibinfo  {journal} {Annals of Physics}\ }\textbf {\bibinfo
			{volume} {303}},\ \bibinfo {pages} {2} (\bibinfo {year} {2003})}\BibitemShut
	{NoStop}%
	\bibitem [{\citenamefont {Sarma}\ \emph {et~al.}(2015)\citenamefont {Sarma},
		\citenamefont {Freedman},\ and\ \citenamefont {Nayak}}]{sarma2015majorana}%
	\BibitemOpen
	\bibfield  {author} {\bibinfo {author} {\bibfnamefont {S.~D.}\ \bibnamefont
			{Sarma}}, \bibinfo {author} {\bibfnamefont {M.}~\bibnamefont {Freedman}}, \
		and\ \bibinfo {author} {\bibfnamefont {C.}~\bibnamefont {Nayak}},\ }\href
	{\doibase 10.1038/npjqi.2015.1} {\bibfield  {journal} {\bibinfo  {journal}
			{npj Quantum Information}\ }\textbf {\bibinfo {volume} {1}},\ \bibinfo
		{pages} {1} (\bibinfo {year} {2015})}\BibitemShut {NoStop}%
	\bibitem [{\citenamefont {Nayak}\ \emph {et~al.}(2008)\citenamefont {Nayak},
		\citenamefont {Simon}, \citenamefont {Stern}, \citenamefont {Freedman},\ and\
		\citenamefont {Das~Sarma}}]{rmptqc}%
	\BibitemOpen
	\bibfield  {author} {\bibinfo {author} {\bibfnamefont {C.}~\bibnamefont
			{Nayak}}, \bibinfo {author} {\bibfnamefont {S.~H.}\ \bibnamefont {Simon}},
		\bibinfo {author} {\bibfnamefont {A.}~\bibnamefont {Stern}}, \bibinfo
		{author} {\bibfnamefont {M.}~\bibnamefont {Freedman}}, \ and\ \bibinfo
		{author} {\bibfnamefont {S.}~\bibnamefont {Das~Sarma}},\ }\href {\doibase
		10.1103/RevModPhys.80.1083} {\bibfield  {journal} {\bibinfo  {journal} {Rev.
				Mod. Phys.}\ }\textbf {\bibinfo {volume} {80}},\ \bibinfo {pages} {1083}
		(\bibinfo {year} {2008})}\BibitemShut {NoStop}%
	\bibitem [{\citenamefont {Zou}\ \emph {et~al.}(2016)\citenamefont {Zou},
		\citenamefont {Brand}, \citenamefont {Liu},\ and\ \citenamefont
		{Hu}}]{Zou2016}%
	\BibitemOpen
	\bibfield  {author} {\bibinfo {author} {\bibfnamefont {P.}~\bibnamefont
			{Zou}}, \bibinfo {author} {\bibfnamefont {J.}~\bibnamefont {Brand}}, \bibinfo
		{author} {\bibfnamefont {X.-J.}\ \bibnamefont {Liu}}, \ and\ \bibinfo
		{author} {\bibfnamefont {H.}~\bibnamefont {Hu}},\ }\href {\doibase
		10.1103/PhysRevLett.117.225302} {\bibfield  {journal} {\bibinfo  {journal}
			{Phys. Rev. Lett.}\ }\textbf {\bibinfo {volume} {117}},\ \bibinfo {pages}
		{225302} (\bibinfo {year} {2016})}\BibitemShut {NoStop}%
	\bibitem [{\citenamefont {Xu}\ \emph {et~al.}(2014)\citenamefont {Xu},
		\citenamefont {Mao}, \citenamefont {Wu},\ and\ \citenamefont
		{Zhang}}]{Xu2014}%
	\BibitemOpen
	\bibfield  {author} {\bibinfo {author} {\bibfnamefont {Y.}~\bibnamefont
			{Xu}}, \bibinfo {author} {\bibfnamefont {L.}~\bibnamefont {Mao}}, \bibinfo
		{author} {\bibfnamefont {B.}~\bibnamefont {Wu}}, \ and\ \bibinfo {author}
		{\bibfnamefont {C.}~\bibnamefont {Zhang}},\ }\href {\doibase
		10.1103/PhysRevLett.113.130404} {\bibfield  {journal} {\bibinfo  {journal}
			{Phys. Rev. Lett.}\ }\textbf {\bibinfo {volume} {113}},\ \bibinfo {pages}
		{130404} (\bibinfo {year} {2014})}\BibitemShut {NoStop}%
	\bibitem [{\citenamefont {Liu}(2015)}]{Liu2015}%
	\BibitemOpen
	\bibfield  {author} {\bibinfo {author} {\bibfnamefont {X.-J.}\ \bibnamefont
			{Liu}},\ }\href {\doibase 10.1103/PhysRevA.91.023610} {\bibfield  {journal}
		{\bibinfo  {journal} {Phys. Rev. A}\ }\textbf {\bibinfo {volume} {91}},\
		\bibinfo {pages} {023610} (\bibinfo {year} {2015})}\BibitemShut {NoStop}%
	\bibitem [{\citenamefont {Thompson}\ \emph {et~al.}(2020)\citenamefont
		{Thompson}, \citenamefont {Brand},\ and\ \citenamefont
		{Z\"ulicke}}]{Thompson2020}%
	\BibitemOpen
	\bibfield  {author} {\bibinfo {author} {\bibfnamefont {K.}~\bibnamefont
			{Thompson}}, \bibinfo {author} {\bibfnamefont {J.}~\bibnamefont {Brand}}, \
		and\ \bibinfo {author} {\bibfnamefont {U.}~\bibnamefont {Z\"ulicke}},\ }\href
	{\doibase 10.1103/PhysRevA.101.013613} {\bibfield  {journal} {\bibinfo
			{journal} {Phys. Rev. A}\ }\textbf {\bibinfo {volume} {101}},\ \bibinfo
		{pages} {013613} (\bibinfo {year} {2020})}\BibitemShut {NoStop}%
	\bibitem [{foo({\natexlab{a}})}]{footnoteMZM}%
	\BibitemOpen
	\href@noop {} {\emph {\bibinfo {title} {\rm{Note that our Hamiltonian is
					different from the one adopted in Refs.~\cite{LiuHu2012,Liu2015} denoted as
					$\tilde{H}$. There exists a unitary transformation $U$, such that
					$U^{\dagger} H U=\tilde{H}$, where $U= \left[ {\begin{array}{cc} T & 0 \\ 0 &
						i \sigma_x T \\ \end{array} }\right]$, where $T=\frac{1}{\sqrt{2}} \left[
					{\begin{array}{cc} 1 & i \\ 1 & -i \\ \end{array} }\right]$. The
					wavefunctions are related through
					$\tilde{u}_{\uparrow}=(u_{\uparrow}+u_{\downarrow})/\sqrt{2}$,
					$\tilde{u}_{\downarrow}=i(u_{\downarrow}-u_{\uparrow})/\sqrt{2}$ },
				$\tilde{v}_{\uparrow}=-i(v_{\uparrow}+v_{\downarrow})/\sqrt{2}$, and
				$\tilde{v}_{\downarrow}=(v_{\uparrow}-v_{\downarrow})/\sqrt{2}$. Therefore,
				when $H$ gives rise to complex wave functions with a spin-balanced
				population, $\tilde{H}$ can produce real wave functions with spin
				imbalance.}}}\BibitemShut {Stop}%
	\bibitem [{foo({\natexlab{b}})}]{footnoteC}%
	\BibitemOpen
	\href@noop {} {\emph {\bibinfo {title} {\rm{The BdG equation
					Eq.~\eqref{eq:BdG} has the particle-hole symmetry, i.e., $C \psi^{*}_{E_j}=
					\psi_{-E_j}$, where $C=e^{i\phi} \left[ {\begin{array}{cc} 0 & I_2 \\ I_2 & 0
						\\ \end{array} }\right]$, $I_2$ is an identity matrix and $\phi$ is an
					arbitrary global phase, satisfying $C^{*}C=I_4$. The BdG Hamiltonian
					satisfies $-C^{*}HC=H^{*}$. Explicitly, this symmetry implies that
					$u_{\sigma}\rightarrow e^{i \phi }v^{*}_{\sigma}$ as $E_j \rightarrow
					-E_j$}}}}\BibitemShut {NoStop}%
	\bibitem [{\citenamefont {Chamon}\ \emph {et~al.}(2010)\citenamefont {Chamon},
		\citenamefont {Jackiw}, \citenamefont {Nishida}, \citenamefont {Pi},\ and\
		\citenamefont {Santos}}]{Chamon2010}%
	\BibitemOpen
	\bibfield  {author} {\bibinfo {author} {\bibfnamefont {C.}~\bibnamefont
			{Chamon}}, \bibinfo {author} {\bibfnamefont {R.}~\bibnamefont {Jackiw}},
		\bibinfo {author} {\bibfnamefont {Y.}~\bibnamefont {Nishida}}, \bibinfo
		{author} {\bibfnamefont {S.-Y.}\ \bibnamefont {Pi}}, \ and\ \bibinfo {author}
		{\bibfnamefont {L.}~\bibnamefont {Santos}},\ }\href {\doibase
		10.1103/PhysRevB.81.224515} {\bibfield  {journal} {\bibinfo  {journal} {Phys.
				Rev. B}\ }\textbf {\bibinfo {volume} {81}},\ \bibinfo {pages} {224515}
		(\bibinfo {year} {2010})}\BibitemShut {NoStop}%
	\bibitem [{\citenamefont {Jackiw}(2012)}]{Jackiw_2012}%
	\BibitemOpen
	\bibfield  {author} {\bibinfo {author} {\bibfnamefont {R.}~\bibnamefont
			{Jackiw}},\ }\href {\doibase 10.1088/0031-8949/2012/t146/014005} {\bibfield
		{journal} {\bibinfo  {journal} {Physica Scripta}\ }\textbf {\bibinfo {volume}
			{T146}},\ \bibinfo {pages} {014005} (\bibinfo {year} {2012})}\BibitemShut
	{NoStop}%
	\bibitem [{foo({\natexlab{c}})}]{footnoteFS}%
	\BibitemOpen
	\href@noop {} {\emph {\bibinfo {title} {\rm{This prescription applies
					whenever the momentum distribution shows sharp, step-like changes separating
					ranges of states with different, roughly constant population }}}}\BibitemShut
	{NoStop}%
	\bibitem [{\citenamefont {de~Gennes}(2018)}]{deGennes}%
	\BibitemOpen
	\bibfield  {author} {\bibinfo {author} {\bibfnamefont {P.~G.}\ \bibnamefont
			{de~Gennes}},\ }\href@noop {} {\emph {\bibinfo {title} {Superconductivity of
				metals and alloys}}}\ (\bibinfo  {publisher} {CRC press},\ \bibinfo {address}
	{Boca Raton},\ \bibinfo {year} {2018})\BibitemShut {NoStop}%
	\bibitem [{foo({\natexlab{d}})}]{footnoteHW}%
	\BibitemOpen
	\href@noop {} {\emph {\bibinfo {title} {\rm{Comparing to a harmonic trap, the
					system in a flat-bottom trap is cleaner and the so-called partial topological
					superfluid~\cite{LiuHu2012,Xu2014} is avoided. We use a a flat-bottom,
					asymmetric trap $V_{\rm ext}(x)= V_0\,[2 + \tanh(1.6 k_\mu x - 0.72\,k_\mu
					L)- \tanh(0.53 k_\mu x + 0.24\,k_\mu L)]$, where L is the computational-box
					length, and $V_0\approx 14 \mu$ }}}}\BibitemShut {NoStop}%
	\bibitem [{\citenamefont {Johnson}(1988)}]{Johnson1988}%
	\BibitemOpen
	\bibfield  {author} {\bibinfo {author} {\bibfnamefont {D.~D.}\ \bibnamefont
			{Johnson}},\ }\href {\doibase 10.1103/PhysRevB.38.12807} {\bibfield
		{journal} {\bibinfo  {journal} {Phys. Rev. B}\ }\textbf {\bibinfo {volume}
			{38}},\ \bibinfo {pages} {12807} (\bibinfo {year} {1988})}\BibitemShut
	{NoStop}%
	\bibitem [{foo({\natexlab{e}})}]{footenoteABS}%
	\BibitemOpen
	\href@noop {} {\emph {\bibinfo {title} {\rm{We adopted a similar criteria to
					identify ABSs used in Refs.~\cite{Liu2015} }.}}}\BibitemShut {Stop}%
	\bibitem [{foo({\natexlab{f}})}]{footnoteII}%
	\BibitemOpen
	\href@noop {} {\emph {\bibinfo {title} {\rm{Let us denote the separation
					between the two Fermi surfaces as $\Delta k_{\rm F}=k_{F+}-k_{F-}$ and the
					width of the condensation peaks $F_{k}$ as $\delta k$. There exists a
					threshold linear coupling $\nu_*$, beyond which, $\nu >\nu_*$, the separation
					of the two Fermi surfaces is large enough for the condensation peaks to be
					independently resolved, i.e., $\Delta k_{\rm F} \gg \delta k \sim
					|\Delta_\infty| m/\hbar^2 k_{F_{+}}=1/\xi_{+}$~\cite{deGennes}
					(Fig.\ref{fig:FGso}), and hence the condition $\Delta k_{\rm F} \xi_+ \gg 1 $
					to support the existence of type-II solitons.}}}}\BibitemShut {Stop}%
	\bibitem [{\citenamefont {Kells}\ \emph {et~al.}(2012)\citenamefont {Kells},
		\citenamefont {Meidan},\ and\ \citenamefont {Brouwer}}]{Brouwer2012}%
	\BibitemOpen
	\bibfield  {author} {\bibinfo {author} {\bibfnamefont {G.}~\bibnamefont
			{Kells}}, \bibinfo {author} {\bibfnamefont {D.}~\bibnamefont {Meidan}}, \
		and\ \bibinfo {author} {\bibfnamefont {P.~W.}\ \bibnamefont {Brouwer}},\
	}\href {\doibase 10.1103/PhysRevB.86.100503} {\bibfield  {journal} {\bibinfo
			{journal} {Phys. Rev. B}\ }\textbf {\bibinfo {volume} {86}},\ \bibinfo
		{pages} {100503} (\bibinfo {year} {2012})}\BibitemShut {NoStop}%
	\bibitem [{\citenamefont {Cayao}\ \emph {et~al.}(2015)\citenamefont {Cayao},
		\citenamefont {Prada}, \citenamefont {San-Jose},\ and\ \citenamefont
		{Aguado}}]{Jorge2015}%
	\BibitemOpen
	\bibfield  {author} {\bibinfo {author} {\bibfnamefont {J.}~\bibnamefont
			{Cayao}}, \bibinfo {author} {\bibfnamefont {E.}~\bibnamefont {Prada}},
		\bibinfo {author} {\bibfnamefont {P.}~\bibnamefont {San-Jose}}, \ and\
		\bibinfo {author} {\bibfnamefont {R.}~\bibnamefont {Aguado}},\ }\href
	{\doibase 10.1103/PhysRevB.91.024514} {\bibfield  {journal} {\bibinfo
			{journal} {Phys. Rev. B}\ }\textbf {\bibinfo {volume} {91}},\ \bibinfo
		{pages} {024514} (\bibinfo {year} {2015})}\BibitemShut {NoStop}%
	\bibitem [{\citenamefont {Liu}\ \emph {et~al.}(2017)\citenamefont {Liu},
		\citenamefont {Sau}, \citenamefont {Stanescu},\ and\ \citenamefont
		{Das~Sarma}}]{Liu2017}%
	\BibitemOpen
	\bibfield  {author} {\bibinfo {author} {\bibfnamefont {C.-X.}\ \bibnamefont
			{Liu}}, \bibinfo {author} {\bibfnamefont {J.~D.}\ \bibnamefont {Sau}},
		\bibinfo {author} {\bibfnamefont {T.~D.}\ \bibnamefont {Stanescu}}, \ and\
		\bibinfo {author} {\bibfnamefont {S.}~\bibnamefont {Das~Sarma}},\ }\href
	{\doibase 10.1103/PhysRevB.96.075161} {\bibfield  {journal} {\bibinfo
			{journal} {Phys. Rev. B}\ }\textbf {\bibinfo {volume} {96}},\ \bibinfo
		{pages} {075161} (\bibinfo {year} {2017})}\BibitemShut {NoStop}%
	\bibitem [{\citenamefont {Fleckenstein}\ \emph {et~al.}(2018)\citenamefont
		{Fleckenstein}, \citenamefont {Dom\'{\i}nguez}, \citenamefont
		{Traverso~Ziani},\ and\ \citenamefont {Trauzettel}}]{Fleckenstein2018}%
	\BibitemOpen
	\bibfield  {author} {\bibinfo {author} {\bibfnamefont {C.}~\bibnamefont
			{Fleckenstein}}, \bibinfo {author} {\bibfnamefont {F.}~\bibnamefont
			{Dom\'{\i}nguez}}, \bibinfo {author} {\bibfnamefont {N.}~\bibnamefont
			{Traverso~Ziani}}, \ and\ \bibinfo {author} {\bibfnamefont {B.}~\bibnamefont
			{Trauzettel}},\ }\href {\doibase 10.1103/PhysRevB.97.155425} {\bibfield
		{journal} {\bibinfo  {journal} {Phys. Rev. B}\ }\textbf {\bibinfo {volume}
			{97}},\ \bibinfo {pages} {155425} (\bibinfo {year} {2018})}\BibitemShut
	{NoStop}%
	\bibitem [{\citenamefont {Moore}\ \emph {et~al.}(2018)\citenamefont {Moore},
		\citenamefont {Stanescu},\ and\ \citenamefont {Tewari}}]{Moore2018}%
	\BibitemOpen
	\bibfield  {author} {\bibinfo {author} {\bibfnamefont {C.}~\bibnamefont
			{Moore}}, \bibinfo {author} {\bibfnamefont {T.~D.}\ \bibnamefont {Stanescu}},
		\ and\ \bibinfo {author} {\bibfnamefont {S.}~\bibnamefont {Tewari}},\ }\href
	{\doibase 10.1103/PhysRevB.97.165302} {\bibfield  {journal} {\bibinfo
			{journal} {Phys. Rev. B}\ }\textbf {\bibinfo {volume} {97}},\ \bibinfo
		{pages} {165302} (\bibinfo {year} {2018})}\BibitemShut {NoStop}%
	\bibitem [{\citenamefont {Reeg}\ \emph {et~al.}(2018)\citenamefont {Reeg},
		\citenamefont {Dmytruk}, \citenamefont {Chevallier}, \citenamefont {Loss},\
		and\ \citenamefont {Klinovaja}}]{Christopher2018}%
	\BibitemOpen
	\bibfield  {author} {\bibinfo {author} {\bibfnamefont {C.}~\bibnamefont
			{Reeg}}, \bibinfo {author} {\bibfnamefont {O.}~\bibnamefont {Dmytruk}},
		\bibinfo {author} {\bibfnamefont {D.}~\bibnamefont {Chevallier}}, \bibinfo
		{author} {\bibfnamefont {D.}~\bibnamefont {Loss}}, \ and\ \bibinfo {author}
		{\bibfnamefont {J.}~\bibnamefont {Klinovaja}},\ }\href {\doibase
		10.1103/PhysRevB.98.245407} {\bibfield  {journal} {\bibinfo  {journal} {Phys.
				Rev. B}\ }\textbf {\bibinfo {volume} {98}},\ \bibinfo {pages} {245407}
		(\bibinfo {year} {2018})}\BibitemShut {NoStop}%
	\bibitem [{\citenamefont {Woods}\ \emph {et~al.}(2019)\citenamefont {Woods},
		\citenamefont {Chen}, \citenamefont {Frolov},\ and\ \citenamefont
		{Stanescu}}]{Woods2019}%
	\BibitemOpen
	\bibfield  {author} {\bibinfo {author} {\bibfnamefont {B.~D.}\ \bibnamefont
			{Woods}}, \bibinfo {author} {\bibfnamefont {J.}~\bibnamefont {Chen}},
		\bibinfo {author} {\bibfnamefont {S.~M.}\ \bibnamefont {Frolov}}, \ and\
		\bibinfo {author} {\bibfnamefont {T.~D.}\ \bibnamefont {Stanescu}},\ }\href
	{\doibase 10.1103/PhysRevB.100.125407} {\bibfield  {journal} {\bibinfo
			{journal} {Phys. Rev. B}\ }\textbf {\bibinfo {volume} {100}},\ \bibinfo
		{pages} {125407} (\bibinfo {year} {2019})}\BibitemShut {NoStop}%
	\bibitem [{\citenamefont {Liu}\ \emph {et~al.}(2019)\citenamefont {Liu},
		\citenamefont {Sau}, \citenamefont {Stanescu},\ and\ \citenamefont
		{Das~Sarma}}]{Liu2019}%
	\BibitemOpen
	\bibfield  {author} {\bibinfo {author} {\bibfnamefont {C.-X.}\ \bibnamefont
			{Liu}}, \bibinfo {author} {\bibfnamefont {J.~D.}\ \bibnamefont {Sau}},
		\bibinfo {author} {\bibfnamefont {T.~D.}\ \bibnamefont {Stanescu}}, \ and\
		\bibinfo {author} {\bibfnamefont {S.}~\bibnamefont {Das~Sarma}},\ }\href
	{\doibase 10.1103/PhysRevB.99.024510} {\bibfield  {journal} {\bibinfo
			{journal} {Phys. Rev. B}\ }\textbf {\bibinfo {volume} {99}},\ \bibinfo
		{pages} {024510} (\bibinfo {year} {2019})}\BibitemShut {NoStop}%
	\bibitem [{\citenamefont {Chen}\ \emph {et~al.}(2019)\citenamefont {Chen},
		\citenamefont {Woods}, \citenamefont {Yu}, \citenamefont {Hocevar},
		\citenamefont {Car}, \citenamefont {Plissard}, \citenamefont {Bakkers},
		\citenamefont {Stanescu},\ and\ \citenamefont {Frolov}}]{Chen2019}%
	\BibitemOpen
	\bibfield  {author} {\bibinfo {author} {\bibfnamefont {J.}~\bibnamefont
			{Chen}}, \bibinfo {author} {\bibfnamefont {B.~D.}\ \bibnamefont {Woods}},
		\bibinfo {author} {\bibfnamefont {P.}~\bibnamefont {Yu}}, \bibinfo {author}
		{\bibfnamefont {M.}~\bibnamefont {Hocevar}}, \bibinfo {author} {\bibfnamefont
			{D.}~\bibnamefont {Car}}, \bibinfo {author} {\bibfnamefont {S.~R.}\
			\bibnamefont {Plissard}}, \bibinfo {author} {\bibfnamefont {E.~P. A.~M.}\
			\bibnamefont {Bakkers}}, \bibinfo {author} {\bibfnamefont {T.~D.}\
			\bibnamefont {Stanescu}}, \ and\ \bibinfo {author} {\bibfnamefont {S.~M.}\
			\bibnamefont {Frolov}},\ }\href {\doibase 10.1103/PhysRevLett.123.107703}
	{\bibfield  {journal} {\bibinfo  {journal} {Phys. Rev. Lett.}\ }\textbf
		{\bibinfo {volume} {123}},\ \bibinfo {pages} {107703} (\bibinfo {year}
		{2019})}\BibitemShut {NoStop}%
	\bibitem [{\citenamefont {J\"unger}\ \emph {et~al.}(2020)\citenamefont
		{J\"unger}, \citenamefont {Delagrange}, \citenamefont {Chevallier},
		\citenamefont {Lehmann}, \citenamefont {Dick}, \citenamefont {Thelander},
		\citenamefont {Klinovaja}, \citenamefont {Loss}, \citenamefont
		{Baumgartner},\ and\ \citenamefont {Sch\"onenberger}}]{Christian2020}%
	\BibitemOpen
	\bibfield  {author} {\bibinfo {author} {\bibfnamefont {C.}~\bibnamefont
			{J\"unger}}, \bibinfo {author} {\bibfnamefont {R.}~\bibnamefont
			{Delagrange}}, \bibinfo {author} {\bibfnamefont {D.}~\bibnamefont
			{Chevallier}}, \bibinfo {author} {\bibfnamefont {S.}~\bibnamefont {Lehmann}},
		\bibinfo {author} {\bibfnamefont {K.~A.}\ \bibnamefont {Dick}}, \bibinfo
		{author} {\bibfnamefont {C.}~\bibnamefont {Thelander}}, \bibinfo {author}
		{\bibfnamefont {J.}~\bibnamefont {Klinovaja}}, \bibinfo {author}
		{\bibfnamefont {D.}~\bibnamefont {Loss}}, \bibinfo {author} {\bibfnamefont
			{A.}~\bibnamefont {Baumgartner}}, \ and\ \bibinfo {author} {\bibfnamefont
			{C.}~\bibnamefont {Sch\"onenberger}},\ }\href {\doibase
		10.1103/PhysRevLett.125.017701} {\bibfield  {journal} {\bibinfo  {journal}
			{Phys. Rev. Lett.}\ }\textbf {\bibinfo {volume} {125}},\ \bibinfo {pages}
		{017701} (\bibinfo {year} {2020})}\BibitemShut {NoStop}%
	\bibitem [{\citenamefont {Fan}\ \emph {et~al.}(2019)\citenamefont {Fan},
		\citenamefont {Zhang}, \citenamefont {Ren},\ and\ \citenamefont
		{Xu}}]{Fan2019}%
	\BibitemOpen
	\bibfield  {author} {\bibinfo {author} {\bibfnamefont {X.}~\bibnamefont
			{Fan}}, \bibinfo {author} {\bibfnamefont {X.}~\bibnamefont {Zhang}}, \bibinfo
		{author} {\bibfnamefont {Z.}~\bibnamefont {Ren}}, \ and\ \bibinfo {author}
		{\bibfnamefont {C.}~\bibnamefont {Xu}},\ }\href {\doibase
		10.1103/PhysRevA.99.013612} {\bibfield  {journal} {\bibinfo  {journal} {Phys.
				Rev. A}\ }\textbf {\bibinfo {volume} {99}},\ \bibinfo {pages} {013612}
		(\bibinfo {year} {2019})}\BibitemShut {NoStop}%
	\bibitem [{\citenamefont {Sau}\ and\ \citenamefont {Tewari}(2021)}]{Sau2021}%
	\BibitemOpen
	\bibfield  {author} {\bibinfo {author} {\bibfnamefont {J.}~\bibnamefont
			{Sau}}\ and\ \bibinfo {author} {\bibfnamefont {S.}~\bibnamefont {Tewari}},\
	}\href {https://arxiv.org/abs/2105.03769} {\bibfield  {journal} {\bibinfo
			{journal} {arXiv preprint arXiv:2105.03769}\ } (\bibinfo {year}
		{2021})}\BibitemShut {NoStop}%
	\bibitem [{\citenamefont {Kong}\ \emph {et~al.}(2021)\citenamefont {Kong},
		\citenamefont {Fan}, \citenamefont {Peng}, \citenamefont {Chen},
		\citenamefont {Zhao},\ and\ \citenamefont {Zou}}]{Kong2021}%
	\BibitemOpen
	\bibfield  {author} {\bibinfo {author} {\bibfnamefont {L.}~\bibnamefont
			{Kong}}, \bibinfo {author} {\bibfnamefont {G.}~\bibnamefont {Fan}}, \bibinfo
		{author} {\bibfnamefont {S.-G.}\ \bibnamefont {Peng}}, \bibinfo {author}
		{\bibfnamefont {X.-L.}\ \bibnamefont {Chen}}, \bibinfo {author}
		{\bibfnamefont {H.}~\bibnamefont {Zhao}}, \ and\ \bibinfo {author}
		{\bibfnamefont {P.}~\bibnamefont {Zou}},\ }\href {\doibase
		10.1103/PhysRevA.103.063318} {\bibfield  {journal} {\bibinfo  {journal}
			{Phys. Rev. A}\ }\textbf {\bibinfo {volume} {103}},\ \bibinfo {pages}
		{063318} (\bibinfo {year} {2021})}\BibitemShut {NoStop}%
	\bibitem [{\citenamefont {Fritsch}\ \emph {et~al.}(2020)\citenamefont
		{Fritsch}, \citenamefont {Lu}, \citenamefont {Reid}, \citenamefont
		{Pi\~neiro},\ and\ \citenamefont {Spielman}}]{Fritsch2020}%
	\BibitemOpen
	\bibfield  {author} {\bibinfo {author} {\bibfnamefont {A.~R.}\ \bibnamefont
			{Fritsch}}, \bibinfo {author} {\bibfnamefont {M.}~\bibnamefont {Lu}},
		\bibinfo {author} {\bibfnamefont {G.~H.}\ \bibnamefont {Reid}}, \bibinfo
		{author} {\bibfnamefont {A.~M.}\ \bibnamefont {Pi\~neiro}}, \ and\ \bibinfo
		{author} {\bibfnamefont {I.~B.}\ \bibnamefont {Spielman}},\ }\href {\doibase
		10.1103/PhysRevA.101.053629} {\bibfield  {journal} {\bibinfo  {journal}
			{Phys. Rev. A}\ }\textbf {\bibinfo {volume} {101}},\ \bibinfo {pages}
		{053629} (\bibinfo {year} {2020})}\BibitemShut {NoStop}%
	\bibitem [{\citenamefont {Schirotzek}\ \emph {et~al.}(2008)\citenamefont
		{Schirotzek}, \citenamefont {Shin}, \citenamefont {Schunck},\ and\
		\citenamefont {Ketterle}}]{Schirotzek2008}%
	\BibitemOpen
	\bibfield  {author} {\bibinfo {author} {\bibfnamefont {A.}~\bibnamefont
			{Schirotzek}}, \bibinfo {author} {\bibfnamefont {Y.-i.}\ \bibnamefont
			{Shin}}, \bibinfo {author} {\bibfnamefont {C.~H.}\ \bibnamefont {Schunck}}, \
		and\ \bibinfo {author} {\bibfnamefont {W.}~\bibnamefont {Ketterle}},\ }\href
	{\doibase 10.1103/PhysRevLett.101.140403} {\bibfield  {journal} {\bibinfo
			{journal} {Phys. Rev. Lett.}\ }\textbf {\bibinfo {volume} {101}},\ \bibinfo
		{pages} {140403} (\bibinfo {year} {2008})}\BibitemShut {NoStop}%
	\bibitem [{\citenamefont {Asano}\ \emph {et~al.}(2006)\citenamefont {Asano},
		\citenamefont {Tanaka},\ and\ \citenamefont {Kashiwaya}}]{Asano2006}%
	\BibitemOpen
	\bibfield  {author} {\bibinfo {author} {\bibfnamefont {Y.}~\bibnamefont
			{Asano}}, \bibinfo {author} {\bibfnamefont {Y.}~\bibnamefont {Tanaka}}, \
		and\ \bibinfo {author} {\bibfnamefont {S.}~\bibnamefont {Kashiwaya}},\ }\href
	{\doibase 10.1103/PhysRevLett.96.097007} {\bibfield  {journal} {\bibinfo
			{journal} {Phys. Rev. Lett.}\ }\textbf {\bibinfo {volume} {96}},\ \bibinfo
		{pages} {097007} (\bibinfo {year} {2006})}\BibitemShut {NoStop}%
\end{thebibliography}
\bibliographystyle{apsrev4-1}

\end{document}